\documentclass{article}
\usepackage{iclr2027_conference,times}

\usepackage{amsmath}
\usepackage{amssymb}
\usepackage{booktabs}
\usepackage{graphicx}
\usepackage{multirow}
\usepackage{url}
\usepackage[T1]{fontenc}

\newcommand{\sai}{s_{\mathrm{ai\text{-}base}}}
\newcommand{\shum}{s_{\mathrm{base\text{-}hum}}}
\newcommand{\slr}{s_{\mathrm{ai\text{-}hum}}}
\newcommand{\tbd}{n/r}
\newcommand{\ci}[3]{$#1$\,[$#2$,$#3$]}
\newcommand{\cis}[3]{$#1^{\ast}$\,[$#2$,$#3$]}
\newcommand{\Nhuman}{2{,}541}
\newcommand{\tabconv}{Metrics, negative pool, chance levels and bolding follow
Section~\ref{sec:setup:metrics}.}

\title{From Tokens to Files: Multi-Granularity Detection and Localization of
AI-Involved Code}

\author{%
\begin{tabular}[t]{@{}p{0.45\textwidth}p{0.45\textwidth}@{}}
Yue Xu & Xiyang Wang \\
\mdseries Peking University & \mdseries Baidu Inc. \\
\mdseries\texttt{yxu25@stu.pku.edu.cn} & \mdseries\texttt{wangxiyang01@baidu.com} \\[3ex]
Hui Liu & Bolei He\thanks{Corresponding author.} \\
\mdseries Baidu Inc. & \mdseries Baidu Inc. \\
\mdseries\texttt{liuhui40@baidu.com} & \mdseries\texttt{hebolei@baidu.com}
\end{tabular}}

\iclrfinalcopy

\begin{document}

\maketitle
\lhead{Preprint}

\begin{abstract}
The increasing use of large language models for code generation has created an
urgent need for reliable methods to distinguish purely human-written code from
\emph{AI-involved} code --- files in which any part was generated or
substantially modified by a model, however small that part is. Our goal is not
only to determine whether a program is purely human-written or has been
generated or modified by AI, but also to localize the specific lines or blocks
attributable to AI. Existing approaches typically compress a file into a single
score, either from surface-level stylistic features or from a global code
representation, which averages away the evidence when only a few blocks of a
file are AI-involved. We propose a detection framework that models AI
authorship at the token level and supports predictions at the line, block, and
file levels. A frozen code language model is equipped with two low-rank
adapters, trained separately on machine-written and human-written files. The
per-token log-likelihood ratio between the two adapters provides local evidence
of AI authorship. Lightweight prediction heads then aggregate this evidence
within lines and blocks, using the surrounding code context to estimate the
probability that each line or block was generated or substantially modified by
AI. The resulting local predictions are further aggregated to produce a
file-level decision. We train these predictions using mixed-authorship files
constructed from human originals and their AI-generated or AI-modified
counterparts, including continuation and fill-in-the-middle scenarios. On an
internal test set of $11{,}773$ AI-involved files across six generation
regimes against a shared pool of $\Nhuman$ human files, our method reaches
$72.4$ macro F1 at a $1\%$ FPR operating point. With no adaptation at
all, the method reaches $62.5$ line-level F1 on HybridCodeAuthorship
\citep{hybridcodeauthorship}, an external human--AI co-writing benchmark that
ships no training split, above the $56$ that the benchmark's own paper reports
as its best line-level result. We further run an extra transfer evaluation on a
human-written-only test set: across $19{,}445$ files from repositories created
before 2022, our false-positive rate stays between $0.55$ and $0.81\%$ when the
false-positive rate threshold is transported at $1\%$. So the operating point
remains valid after transferring from paired synthetic negatives to real human
code.
\end{abstract}

\section{Introduction}
\label{sec:intro}

Assistants that write code inside an existing file have moved the unit of
authorship below the file. A commit may contain a completed function body, a
refilled interior span, or a locally refactored block, with the rest of the file
untouched. The question a provenance tool has to answer is therefore not only
\emph{was this file written by a model} but \emph{is any part of it}, and, if a
reviewer is to act on the answer, \emph{which part}.

Two families of detectors dominate the current literature, and both answer only
the first question. Zero-shot likelihood statistics score a file against a
reference language model and threshold a scalar summary: log-rank
\citep{gehrmann2019gltr}, LRR \citep{su2023detectllm}, perturbation curvature
\citep{mitchell2023detectgpt,bao2024fastdetectgpt}, or cross-perplexity between
two models \citep{hans2024binoculars}. Supervised detectors fine-tune a code
encoder \citep{feng2020codebert,guo2022unixcoder,wang2023codet5plus} or align a
proxy model to machine style \citep{song2025imbd} and read a single decision out
of a global representation. In both cases the file is compressed to one number
before any decision is made, which is adequate when the whole file is machine
written and lossy when only a few blocks are.

Mixed authorship breaks these detectors in two distinct ways. A global score
averages a short AI span together with the human majority of the file, so the
evidence is diluted in proportion to the AI fraction. Independently, the span
itself is weaker evidence than a whole machine-written file, because the model
that produced it was conditioned on the surrounding human code and inherited its
naming and layout. Our measurements show both effects: on the two
interior-segment test cells, the strongest supervised file-level baseline drops to
$0.647$ and $0.565$ AUROC, while the same detector reaches
$0.945$--$0.963$ on the from-scratch cells.

We therefore keep the evidence local for as long as possible. A single frozen
code language model is equipped with two LoRA adapters
\citep{hu2022lora}, trained with a plain causal language modeling objective on
machine-written and on human-written files respectively; neither has a
classification head. Their per-token log-likelihood ratio is a local score whose
human-side term cancels the confound of how predictable a token is in the first
place.

Evaluating this design requires separating what a detector cannot represent from
what its training distribution never contained. The transfer--adaptation gap on
external benchmarks is the sharpest finding in the paper, and we read it as a
statement about training corpora rather than detector capacity.

The main contributions of this paper are:
\begin{itemize}
  \item A detection framework that couples a token-level likelihood ratio with
        line- and block-level supervised heads over a single frozen code language
        model, so that local generation evidence and its surrounding context are
        modeled jointly rather than as alternatives.
  \item A mixed-authorship data construction protocol that synthesizes
        continuation and fill-in-the-middle edits with character-level ground
        truth and a controlled AI ratio $\rho$, which supplies within-file
        supervision for the line and block heads and test cells covering both
        tail and interior segment geometries. Appendix~\ref{app:ablation-extra}
        ablates the training-side composition and finds corpus volume, rather
        than the mixture of forms, to be the load-bearing factor.
  \item An evaluation protocol that separates \emph{capacity} from \emph{distribution}
        on external benchmarks under two protocols (zero-adaptation transfer and
        per-benchmark adaptation), reporting results at the file, block and line
        granularities, together with a carrier-artifact control and a
        false-positive measurement on real repository code.
\end{itemize}

We defer a full discussion of related work to Appendix~\ref{app:related}.

\section{Problem Formulation}
\label{sec:problem}

Let $x = (x_1,\dots,x_T)$ be the token sequence of a source file. The file-level
label is binary and \emph{AI-involved}: $y=1$ if any part of the file was
generated by a language model, and $y=0$ if the file is entirely human-written.
Mixed files therefore carry $y=1$ regardless of how little of the file is
machine-written, which is the convention used in machine-revised-text detection
\citep{song2025imbd} and by the \emph{machine-refined} class of DroidCollection
\citep{droidcollection}. A detector scores the file and thresholds that score at
$\tau$, calibrated on human files at a target false-positive rate. The detector
sees the file alone at test time: no repository metadata, commit history or
generation-side information is available.

Ground truth for a mixed file is a set of character intervals covering its
machine-written text. A line is an \emph{AI line} if any of its characters falls
inside one of those intervals. The AI ratio $\rho$ is the
fraction of AI lines in the file, over all lines including blank ones; our
synthesized mixed files carry $\rho \ge 0.05$ by construction
(Section~\ref{sec:method:data}) and pure AI files have $\rho = 1$. A \emph{block}
is a run of $K$ consecutive lines with $K \in \{8,16\}$, cut from the first line
of the file; the last block keeps whatever fewer than $K$ lines remain, since
dropping it would exclude the end of the file from prediction and the
machine-written tail of continuation files sits exactly there. Block $b$ carries
the label $z_b = 1$ if it contains at least one AI line and $z_b = 0$ otherwise,
so a block whose AI lines cover only part of it is positive while being mixed in
content. Blocks and lines are
the two prediction units we report. Every line-level number in the paper is
computed against these line labels, and every block-level number against these
block labels except where a caption states the convention it departs to.

\section{Method}
\label{sec:method}

\subsection{Token-level likelihood ratio}
\label{sec:method:lr}

We keep a single code language model $p_{\theta}$ frozen and train two low-rank
adapters \citep{hu2022lora} on it with a plain causal language modeling
objective: an AI-side adapter on machine-written files and a human-side adapter
on human-written files. Neither adapter has a classification head. The per-token
score is the log-likelihood ratio
\begin{equation}
  s_t \;=\; \log p_{\mathrm{ai}}(x_t \mid x_{<t}) \;-\; \log p_{\mathrm{hum}}(x_t \mid x_{<t}),
  \label{eq:lr}
\end{equation}
which we decompose through the frozen base model as
$\slr = \sai + \shum$, where $\sai = \log p_{\mathrm{ai}} - \log p_{\theta}$ and
$\shum = \log p_{\theta} - \log p_{\mathrm{hum}}$ are the two adapters' deviations
from the shared base model. Neither half is sufficient on its own: subtracting the
human-side term cancels the confound of \emph{how predictable the text is in the
first place}, which a single-sided score cannot separate from authorship
(Section~\ref{sec:ablation}).

Both adapters are rank-$16$ LoRA modules with $\alpha = 32$
($18.5$M trainable parameters per side, $36.9$M in total), trained for one epoch
in fp16 on the pure-file split, with no packing and token-budget batching. The
frozen backbone is Qwen2.5-Coder-1.5B \citep{hui2024qwen25coder}.

\subsection{Line- and block-level discriminative heads}
\label{sec:method:block}

Token scores are mapped to lines (a token belongs to the line containing its last
character) and then to fixed-$K$ blocks. Each block is described by $88$ features
in five families: raw statistics of $s_t$, of the base-model profile
(negative log-likelihood, entropy, rank) and of $\mathrm{dctx}$, the gap between
the long- and short-context negative log-likelihood of the same token under each
of the three passes; within-file $z$-scores of those statistics;
neighbouring-block contrasts; positional features; and language
identity (Table~\ref{tab:featlist}).
A sequence head (bidirectional GRU; \citealp{cho2014gru}) reads the block sequence of a file and emits
a per-block AI probability $\hat z_b$. The head is trained only on blocks that
are entirely AI or entirely human, while reading the whole file, mixed blocks
included, as context; at inference it scores every block, and block-level
metrics are computed against $z_b$ over all blocks.

The line head is the same construction one granularity finer: a bidirectional
GRU over the per-token signal sequence, mean-pooled to one logit and one AI
probability $\hat z_\ell$ per line.
Line labels come directly from the AI intervals, so no line has to be
held out of the loss. Both heads consume the same token-level profile, each
signal paired with its within-file $z$-score, and differ in where it is pooled,
the block head into per-block statistics before the sequence model and the line
head only after it, as well as in their feature sets
(Appendix~\ref{app:impl}). Which granularity is preferable depends on the
regime, so we report both: Section~\ref{sec:fine} gives their
behaviour against AI segment length; a syntax-aligned alternative
to the fixed grid and the two heads at equal input are in
Appendix~\ref{app:ablation-extra}.

\subsection{Aggregation to a file-level score}
\label{sec:method:agg}

Each read-out produces a file-level score $f(x)$ by
averaging its own units: the token ratio over the $T$ tokens of the file, the
block head over the logits of the block set $\mathcal{B}$, and the line head over
the logits of the $L$ lines,
\begin{equation}
  f_{\mathrm{lr}}(x) = \frac{1}{T}\sum_{t=1}^{T} s_t,
  \quad
  f_{\mathrm{blk}}(x) = \frac{1}{|\mathcal{B}|}\sum_{b \in \mathcal{B}} \sigma^{-1}(\hat z_b),
  \quad
  f_{\mathrm{lin}}(x) = \frac{1}{L}\sum_{\ell=1}^{L} \sigma^{-1}(\hat z_\ell),
  \label{eq:agg}
\end{equation}
where $\sigma^{-1}$ is the logit function. The mean is the selected form in all
three cases and not the only one we tried: for the token score we compare it
against a sliding-window maximum, a change-point tail mean and the best segment
after binary segmentation \citep{truong2020changepoint,killick2012pelt}, and for
each head against a per-unit maximum and a learned attention aggregator over unit
states. The aggregator and the block size $K$ are selected on the development
split and applied to test unchanged.

Thresholds are calibrated as stated in Section~\ref{sec:setup:metrics}, except
where a table states otherwise: the block and line read-outs use thresholds frozen
on the development split and transported unchanged, both at their own granularity
and, in Section~\ref{sec:results:fpr}, to the pre-2022 repository corpus, which is
disjoint from the internal paired \textsc{human} cell.

\subsection{Constructing mixed-authorship supervision}
\label{sec:method:data}

Supervision at either granularity requires files whose authorship is
heterogeneous \emph{and} known. We synthesize two forms. In \textbf{continuation} a model completes a file
from its first $N\%$ of lines, which gives exactly one AI segment at the tail. In
\textbf{fill-in-the-middle} one or more spans are removed and refilled, which
gives interior segments \citep{bavarian2022fim}; the number of holes is sampled from $\{1,2,3,5,8\}$.
Sampling is constrained so that $\rho$ is balanced over five buckets covering
$[0.05, 1.0]$, with the cut point or the hole coverage solved backwards from the
target $\rho$ and resampled when the realized ratio falls outside its bucket. AI
spans are recorded as exact whole-line character intervals. For every mixed file
we additionally keep its original human version as a negative example, so the
heads see "no boundary anywhere" files that are otherwise textually paired
with positives.

Two properties of this construction bound what the heads can learn. Both forms
produce long segments --- the median AI segment in the training and development
splits is $20$ non-blank lines --- so the scattered short-segment regime that
in-place refactoring produces is absent from training and appears only on the test
side, where we read it as out-of-distribution measurement rather than as a held-out
sample of the training regime. The dual-LoRA
backbone, by contrast, is trained on pure files only. The development split used
for aggregation and threshold selection does contain mixed files, because the
score of a file moves with its AI ratio $\rho$:
pure files sit at the extremes, one mode per class, while a partially
machine-written file falls between them. A development split of pure files alone
would fix the aggregator choice at scores the mixed part of the test set almost
never produces.

\section{Experimental Setup}
\label{sec:setup}

\subsection{Internal dataset}
\label{sec:setup:data}

Files are drawn from public repositories and paired with machine-written
counterparts produced by a two-stage protocol in which a reader model summarizes
a file into a specification and a writer model reimplements the file from the
specification alone, without seeing the original. The isolation is deliberate:
a writer that sees the original produces \emph{a rewrite}, and a detector trained
on rewrites learns rewriting artifacts. The reader is GPT-5.6 decoded at
temperature $0$; the writers are Claude Opus 5, DeepSeek-V4-Flash, GPT-5.6 and
GLM-5.2, all decoded at temperature $0.7$, and every file is assigned to one of
them at random so that each contributes about a quarter of the machine-written
side. Mixing writers dilutes single-model length and comment-density
fingerprints. Splits are cut by repository hash, so no repository appears on
both sides of a split.

Table~\ref{tab:data} (Appendix~\ref{app:impl}) summarizes the splits: the two LoRA
adapters train on $12{,}002$ pure human/AI files, the block head on $1{,}582$ mixed
files ($843$ continuation and $739$ fill-in-the-middle) together with their
$1{,}582$ human originals, and the aggregator, $K$ and decision
thresholds are fixed on $4{,}062$ pure and $398$ mixed development files. The test
set applies one generation axis at a time to a fixed human base --- $11{,}773$
positives over six generation cells against a shared pool of $\Nhuman$ human
negatives.

\subsection{Test cells}
\label{sec:setup:cells}

The test set varies one axis at a time over a fixed human base: \textsc{human}
(the only negative cell), \textsc{baseline} (specification-based rewriting with
the training writer models), \textsc{unseen-models} (same protocol with four
writer models never seen in training: Qwen3-Coder-Plus, Doubao-Seed-2.1-Pro,
Kimi-K2.6 and MiniMax-M2.7), \textsc{full-rewrite} (the model sees the original
and writes its own implementation), \textsc{completion} (a human prefix is
completed), \textsc{refactor} (in-place refactoring, producing many very short AI
segments), and \textsc{fim} (interior spans refilled). The cells differ in how
much human text survives in the output, which we find to be the dominant
difficulty factor.

\subsection{Baselines}
\label{sec:setup:baselines}

Every supervised baseline shares our training split, the same development split
for aggregation and threshold selection, and, where applicable, the same scoring
backbone. \textbf{Group A (zero-shot)} comprises six likelihood-based
statistics: log-rank \citep{gehrmann2019gltr}, LRR \citep{su2023detectllm},
Fast-DetectGPT \citep{bao2024fastdetectgpt}, Binoculars
\citep{hans2024binoculars}, DetectCodeGPT \citep{shi2024detectcodegpt} and the
AIGCode Detector \citep{xu2024aigcode}. \textbf{Group B (supervised)} comprises
a frozen code encoder (CodeT5+ 220m) with an MLP head, fully fine-tuned
CodeBERT \citep{feng2020codebert} and UniXcoder \citep{guo2022unixcoder}, ImBD
\citep{song2025imbd}, DroidDetect \citep{droidcollection}, and --- the critical
control --- \emph{the same backbone with a LoRA classification head}, matched in
backbone, data and optimisation schedule, so that the only difference from our
method is ``classification head versus likelihood ratio''. \textbf{Group C
(ours)} is the token-level likelihood ratio, the block head at $K \in
\{8,16\}$, the line head, and their combinations. Scoring and aggregation
details are in Appendix~\ref{app:tables}.

\subsection{Metrics}
\label{sec:setup:metrics}

The positive class is \emph{AI-involved}, so a mixed file is positive whatever its
$\rho$ (Section~\ref{sec:problem}). We report AUROC, TPR@1\%FPR and macro F1 over
the positive and negative files of each comparison, scored on the file score $f$.
An AUROC below $0.500$ means the score ranks machine files as \emph{more}
human-like than human files. \textbf{TPR@1\%FPR} reads its threshold off the
negatives of the same comparison; on the internal test set that is the
\textsc{human} cell of the test set itself, so the operating point is exact rather
than transported from development. \textbf{Macro F1} averages the per-class F1 of
the AI class and of the human class, the latter scored on the inverted prediction,
which penalizes a detector that calls everything AI-involved; it needs a threshold
and the available calibration data differ by corpus, so each table states its own.
Accuracy appears only in Appendix~\ref{app:acc}: the fraction of mixed files
differs between development and test, so a transported threshold is systematically
mismatched, which is why the three metrics above are either threshold-free or read
at a fixed FPR. Every table in the paper follows the same conventions for metrics,
negative pool, chance levels and bolding, and states only its departures from them;
table conventions, chance levels, per-cell counts and per-row exceptions are in
Appendix~\ref{app:tables}.

\subsection{External benchmarks and transfer protocols}
\label{sec:setup:external}

We evaluate on three external resources, none of which shares our generation
protocol: \textbf{AICD-Bench} \citep{aicdbench}, used as its binary task T1 and
its four-class task T3; \textbf{DroidCollection}
\citep{droidcollection}, a multi-generator, multi-task corpus with four authorship
classes; and
\textbf{HybridCodeAuthorship} (HCA) \citep{hybridcodeauthorship}, the only one of
the three with per-line ground truth. For the multi-class
resources every machine-involved class is mapped to $y=1$. Two protocols are
reported: \textbf{P1} transfer with no adaptation; and \textbf{P2} adaptation on
the benchmark's own training split, which is the protocol under which published
numbers were obtained. HCA has no training split, so only P1 is defined there, and
it is additionally scored at its native line and block granularity.
Group~A and the Group~B baselines are run on the external benchmarks as well.
Composition, file counts, per-slice results and the row
set of each table are in Appendix~\ref{app:external}.

\section{Results}
\label{sec:results}

\subsection{Main comparison}
\label{sec:results:main}

\begin{table}[tbp]
\small
\caption{File-level detection on the internal test set. Per-cell entries
are macro F1 (\%) at the $1\%$ FPR threshold read off the \textsc{human} cell; at
that threshold a score without signal labels almost every file human, and the
all-negative predictor scores $33.3$--$39.8\%$, which is the floor each column has
to clear. \emph{Avg.} is the unweighted mean over the six cells, given for all
three metrics. Group~C rows are cumulative, not independent detectors. Every cell of
this table with AUROC and TPR@1\%FPR as well is Table~\ref{tab:main-full}
(Appendix~\ref{app:tables}).}
\label{tab:main}
\centering
\scriptsize
\renewcommand{\arraystretch}{1.1}
\setlength{\tabcolsep}{4.5pt}
\begin{tabular}{llcccccc@{\hspace{8pt}}c}
\toprule
& Method & \textsc{base} & \textsc{unseen} & \textsc{rewrite} & \textsc{compl.} & \textsc{refac.} & \textsc{fim} & Avg. \\
& & \multicolumn{6}{c}{macro F1 (\%) at the $1\%$ FPR operating point} & AUROC / TPR@1\%FPR / F1 \\
\midrule
\multirow{6}{*}{A}
& log-rank & $33.3$ & $33.2$ & $35.7$ & $39.1$ & $36.5$ & $40.9$ & $0.488$ / $0.4$ / $36.5$ \\
& LRR & $33.4$ & $33.2$ & $35.7$ & $38.9$ & $36.4$ & $40.3$ & $0.481$ / $0.3$ / $36.3$ \\
& Fast-DetectGPT & $33.3$ & $33.3$ & $35.9$ & $41.8$ & $36.6$ & $40.1$ & $0.481$ / $0.8$ / $36.8$ \\
& Binoculars & $33.2$ & $33.2$ & $35.7$ & $39.1$ & $36.7$ & $40.4$ & $0.488$ / $0.4$ / $36.4$ \\
& DetectCodeGPT & $33.3$ & $33.1$ & $35.7$ & $39.1$ & $36.5$ & $40.6$ & $0.518$ / $0.4$ / $36.4$ \\
& AIGCode Detector & $34.1$ & $33.8$ & $33.8$ & $34.3$ & $33.2$ & $34.7$ & $0.475$ / $0.9$ / $34.0$ \\
\midrule
\multirow{6}{*}{B}
& CodeT5+ + MLP & $63.6$ & $61.7$ & $58.2$ & $40.5$ & $43.4$ & $41.1$ & $0.771$ / $17.3$ / $51.4$ \\
& CodeBERT & $72.2$ & $64.9$ & $69.1$ & $40.6$ & $50.6$ & $41.6$ & $0.791$ / $24.4$ / $56.5$ \\
& UniXcoder & $74.7$ & $68.5$ & $72.5$ & $41.8$ & $51.5$ & $41.7$ & $0.817$ / $27.3$ / $58.4$ \\
& ImBD & $57.0$ & $57.8$ & $53.1$ & $40.8$ & $40.8$ & $41.3$ & $0.723$ / $13.4$ / $48.5$ \\
& \emph{Same backbone} + LoRA cls.\ head & $79.3$ & $71.6$ & $75.0$ & $42.2$ & $52.9$ & $41.3$ & $0.808$ / $30.5$ / $60.4$ \\
& DroidDetect & $82.9$ & $80.9$ & $78.6$ & $43.7$ & $54.9$ & $42.2$ & $0.801$ / $36.1$ / $63.8$ \\
\midrule
\multirow{4}{*}{C}
& Ours: token ratio, file mean & $\mathbf{89.1}$ & $89.4$ & $\mathbf{87.4}$ & $50.0$ & $63.1$ & $45.3$ & $0.838$ / $47.1$ / $70.7$ \\
& + block head, $K=8$ & $86.3$ & $90.7$ & $83.3$ & $\mathbf{60.8}$ & $62.9$ & $\mathbf{50.4}$ & $0.852$ / $\mathbf{48.4}$ / $\mathbf{72.4}$ \\
& + block head, $K=16$ & $87.1$ & $\mathbf{91.6}$ & $83.7$ & $58.5$ & $\mathbf{63.8}$ & $48.6$ & $\mathbf{0.853}$ / $\mathbf{48.4}$ / $72.2$ \\
& + line head, file mean & $79.8$ & $82.9$ & $72.4$ & $57.1$ & $54.8$ & $47.1$ & $0.836$ / $37.4$ / $65.7$ \\
\bottomrule
\end{tabular}
\end{table}

Group A sits at the all-negative floor of the operating point: on the three
from-scratch cells (\textsc{base}, \textsc{unseen}, \textsc{rewrite}) the six
likelihood statistics are within $2.1$ points of it, because TPR at $1\%$ FPR is
essentially zero there. This is not a
scoring-direction error --- flipping Binoculars or Fast-DetectGPT does not recover
a usable detector --- it is the assumption ``machine code is more predictable than
human code'' failing on specification-based rewriting, whose writer never sees the
original. Where human text is left in place, the zero-shot statistics leave the
floor only marginally: $38.9$--$41.8$ on \textsc{compl.} (floor $38.6$),
$40.1$--$40.9$ on \textsc{fim} ($39.8$) and $36.4$--$36.7$ on \textsc{refac.}\
($36.4$).

Group B changes the picture completely: every supervised baseline reaches
$53.1$--$82.9$ on the from-scratch cells, against at most $35.9$ for any zero-shot
statistic, so what makes
specification-based rewriting detectable is supervision on this distribution, not
a better likelihood statistic. The
informative comparison is the same-backbone LoRA classification head, which shares
the backbone, the training files, the tokenizer and the optimisation schedule with
our adapters and, at scoring time, the same $4096$-token window with $512$ tokens of
carried context as our own scorer (its file score is the mean over window logits),
so its residual gap to the token ratio
($60.4$ against $70.7$ average macro F1) is attributable to the likelihood-ratio
formulation rather than to the $1.5$B backbone. The gap has two parts. On the
from-scratch cells ($9.8$--$17.8$ points) it is a tail effect: at the same $1\%$
FPR the token ratio recovers more positives. On \textsc{compl.} and \textsc{fim}
it is structural: the classifier stays $3.6$ and $1.5$ points above the
all-negative floor, against $11.4$ and $5.5$ for the token ratio. A windowed head
compresses each window into one decision and averages the windows, so a short
inserted span is averaged away, whereas a token-level score keeps it addressable.

Difficulty tracks how much human text survives in the output: \textsc{rewrite}
($87.4$) is as easy as the specification-based cells ($89.1$, $89.4$), while
\textsc{compl.} and \textsc{fim}, where a human prefix or envelope is left in
place, are the two hardest ($50.0$ and $45.3$). The block head's gain is
concentrated exactly there ($+10.8$ on \textsc{compl.} and $+5.1$ on \textsc{fim}
at $K=8$); on the from-scratch cells it costs $2.8$ on \textsc{base} and $4.1$ on
\textsc{rewrite} while gaining $1.3$ on \textsc{unseen}, which is the trade-off a
localizing read-out makes and the reason we report both. The line head pushes it
further ($+7.1$ on \textsc{compl.}, $+1.8$ on \textsc{fim}, $-6.5$ to $-15.0$ on the
from-scratch cells) and ends $5.0$ points below the token ratio on average: it
detects files with an interior AI segment rather than files as a whole.

\subsection{Localization value and block size}
\label{sec:fine}

The cell-level averages of Table~\ref{tab:main} conflate how much of a file is AI
with how its AI part is laid out, and only the latter bounds what localization can
add. Stratifying the three mixed cells by the AI ratio $\rho$
(Table~\ref{tab:rho}, Appendix~\ref{app:loc}) shows the bound directly: below
$\rho = 0.2$ the token ratio and both block heads lie within $0.8$ macro F1 of each
other ($50.4$--$51.2$), while at $\rho \ge 0.8$ the $K{=}8$ head is $11.9$ points
above the token ratio ($78.4$ against $66.5$). A supervised unit-level head buys
little when the machine-written part of the file is small.

The block size follows the same logic: what matters is whether a block boundary can
fall inside an AI segment. When the median segment is close to $K$, reducing $K$
improves every bucket; when it is far below any usable $K$, most blocks are
mixed and the larger $K$ wins because each block carries more tokens and
less feature noise. The three mixed cells of Table~\ref{tab:main-full} instantiate the
three regimes: on \textsc{fim} (median segment near $K=8$) and \textsc{compl.} (one
segment far longer than either block) the smaller block wins ($50.4$ vs.\ $48.6$
and $60.8$ vs.\ $58.5$ macro F1), as its interior blocks are pure and the finer grid places
the boundary more precisely; on \textsc{refac.} (median two lines, below any usable
$K$) the order reverses ($62.9$ vs.\ $63.8$): no block isolates a segment, so
tokens per block outweigh boundary precision. The line head is weaker still
($54.8$ file-level macro F1, line-level F1 $48.0$, Table~\ref{tab:linecell}): on
\textsc{refac.}\ the signal comes from pooling, not localization.
Tables~\ref{tab:blockcell} and~\ref{tab:main-full} give both $K$ per cell, and
Table~\ref{tab:seglen} per bucket (Appendices~\ref{app:block}, \ref{app:tables},
\ref{app:loc}).

\subsection{Ablations}
\label{sec:ablation}

The ablations answer five questions: whether the ratio form is doing the work,
which token categories carry it, whether any single feature family is
load-bearing, how much of the performance depends on the amount of adapter
training data, and how much of it depends on the backbone being
Qwen2.5-Coder-1.5B. The components sit at two levels, so the panels are read against two
references: the token-level panels against $\slr$ with the file-mean aggregation,
the block-level panel against the $K=16$ head with logit-mean aggregation. Each row
removes or replaces exactly one component.

\begin{table}[tbp]
\small
\caption{Ablations on the internal test set. Each row is a complete
file-level system, and $\Delta$ is its paired-bootstrap difference from the panel's
reference row ($^{\ast}$: the $95\%$ interval excludes zero). Further ablations:
Appendix~\ref{app:ablation-extra}.}
\label{tab:ablation}
\begin{center}
\scriptsize
\setlength{\tabcolsep}{1pt}
\begin{tabular}{lcccccc}
\toprule
& \multicolumn{2}{c}{All cells} & \multicolumn{2}{c}{Mixed cells} & \multicolumn{2}{c}{TPR@1\%FPR} \\
\cmidrule(lr){2-3}\cmidrule(lr){4-5}\cmidrule(lr){6-7}
Variant & AUROC & $\Delta$ & AUROC & $\Delta$ & \% & $\Delta$ \\
\midrule
\multicolumn{7}{l}{\emph{Token-level panel}; reference: $\slr$, file mean} \\
$\slr$, file mean (ref.)   & 0.838 & -- & 0.707 & -- & 47.1 & -- \\
\quad $\sai$ only                  & 0.773 & \cis{-0.065}{-0.074}{-0.057} & 0.646 & \cis{-0.061}{-0.073}{-0.049} & 18.9 & \cis{-28.2}{-30.5}{-24.2} \\
\quad $\shum$ only                 & 0.506 & \cis{-0.331}{-0.344}{-0.318} & 0.497 & \cis{-0.210}{-0.226}{-0.193} & 1.0 & \cis{-46.0}{-47.6}{-42.7} \\
\quad no ratio (AI-side NLL) & 0.522 & \cis{-0.316}{-0.329}{-0.301} & 0.563 & \cis{-0.144}{-0.160}{-0.128} & 0.5 & \cis{-46.6}{-48.1}{-43.4} \\
\midrule
\multicolumn{7}{l}{\emph{Token categories}, parsed subset; reference: all tokens of the subset} \\
all tokens (ref.)          & 0.837 & -- & 0.707 & -- & 47.1 & -- \\
\quad comments removed             & 0.802 & \cis{-0.035}{-0.041}{-0.029} & 0.688 & \cis{-0.018}{-0.027}{-0.010} & 21.4 & \cis{-25.6}{-28.1}{-20.7} \\
\quad code tokens only             & 0.744 & \cis{-0.094}{-0.103}{-0.084} & 0.645 & \cis{-0.062}{-0.075}{-0.048} & 6.4 & \cis{-40.6}{-43.3}{-37.1} \\
\quad comment tokens only          & 0.787 & \cis{-0.051}{-0.057}{-0.044} & 0.638 & \cis{-0.069}{-0.079}{-0.058} & 22.9 & \cis{-24.2}{-30.6}{-17.0} \\
\midrule
\multicolumn{7}{l}{\emph{Block-level panel}; reference: block head, $K{=}16$, BiGRU, logit mean} \\
block head $K{=}16$ (ref.) & 0.853 & -- & 0.749 & -- & 48.4 & -- \\
\quad $K = 8$                      & 0.852 & \ci{-0.001}{-0.004}{+0.003} & 0.761 & \cis{+0.013}{+0.007}{+0.019} & 48.4 & \ci{-0.0}{-2.9}{+4.0} \\
\quad logistic head, no BiGRU & 0.845 & \cis{-0.008}{-0.013}{-0.003} & 0.750 & \ci{+0.001}{-0.006}{+0.009} & 43.8 & \cis{-4.6}{-7.5}{-0.1} \\
\quad boosted-trees head   & 0.847 & \cis{-0.006}{-0.011}{-0.000} & 0.743 & \ci{-0.006}{-0.015}{+0.004} & 41.3 & \cis{-7.1}{-10.4}{-1.9} \\
\quad no raw family        & 0.845 & \cis{-0.008}{-0.012}{-0.004} & 0.725 & \cis{-0.024}{-0.031}{-0.016} & 47.7 & \ci{-0.7}{-3.4}{+3.5} \\
\quad no $z$-score family  & 0.843 & \cis{-0.011}{-0.014}{-0.007} & 0.737 & \cis{-0.012}{-0.017}{-0.006} & 44.1 & \cis{-4.3}{-7.8}{-0.1} \\
\quad no positional family & 0.852 & \ci{-0.001}{-0.002}{+0.001} & 0.746 & \ci{-0.002}{-0.004}{+0.000} & 47.5 & \ci{-0.9}{-2.7}{+2.1} \\
\quad no language family   & 0.856 & \cis{+0.003}{+0.001}{+0.005} & 0.753 & \cis{+0.005}{+0.001}{+0.008} & 47.9 & \ci{-0.5}{-2.1}{+3.1} \\
\midrule
\multicolumn{7}{l}{\emph{Adapter training data}; reference: $\slr$, file mean} \\
\quad $10\%$ of adapter data & 0.783 & \cis{-0.055}{-0.063}{-0.047} & 0.670 & \cis{-0.038}{-0.051}{-0.025} & 27.9 & \cis{-19.1}{-21.7}{-15.2} \\
\midrule
\multicolumn{7}{l}{\emph{Backbone family}; reference: $\slr$, file mean} \\
\quad deepseek-coder-1.3B backbone & 0.826 & \cis{-0.012}{-0.017}{-0.006} & 0.690 & \cis{-0.017}{-0.026}{-0.007} & 43.2 & \cis{-3.9}{-6.4}{-0.3} \\
\bottomrule
\end{tabular}
\end{center}
\end{table}

The token-level panel is the sharpest of the three. Neither half of the
decomposition is sufficient: the AI-side deviation alone reaches $0.773$ and the
human-side deviation alone $0.506$, i.e.\ chance, against $0.838$ for the ratio.
The asymmetry is the point: a term that cannot rank files at all is worth
$0.065$ AUROC and $28.2$ TPR once subtracted, because what it removes
is how predictable the text is: a normalizer, not a
detector. Replacing the ratio by the AI-side adapter's likelihood
($0.522$) is no better, which rules out reading our score as "the
AI-tuned model finds this file predictable".

In the block-level panel no feature family is individually load-bearing: the only
two whose removal is a statistically detectable \emph{cost} ($-0.008$ for raw,
$-0.011$ for $z$-scores) are about an order of magnitude cheaper than either half of
the ratio, the positional family is within noise on both AUROC columns ($-0.001$ and
$-0.002$), and the language family is \emph{negative} --- removing it improves both
AUROC columns slightly but significantly ($+0.003$ and $+0.005$) and leaves the tail
unchanged within noise ($48.4 \to 47.9$). The positional result also holds where a
shortcut would have to appear, on \textsc{compl.}, whose AI segment is always at
the tail: per-cell $\Delta$ (Table~\ref{tab:dropcell}) costs at most $0.005$ AUROC
in any single cell, so the head does not take the positional shortcut.

The third reading concerns the operating point rather than any component. Rows
that are nearly free in threshold-free terms are expensive at $1\%$ FPR:
removing comment tokens costs $0.035$ AUROC but more than half of the TPR
($47.1 \to 21.4$), and a tenth of the adapter data costs $0.055$ AUROC and again
about $40\%$ of the TPR. Scoring comments \emph{alone} retains $0.787$ AUROC, more
than the code-token-only variant ($0.744$), which is itself $0.244$ above chance,
so comments are neither the mechanism nor a negligible part of it. A
low-false-alarm deployment therefore depends on comment tokens and the full
adapter training set --- two components AUROC alone calls unimportant.

Swapping Qwen2.5-Coder-1.5B for deepseek-coder-1.3B \citep{guo2024deepseekcoder} costs $0.012$ AUROC and
$3.9$ TPR --- a fifth of the data-quantity row, reversing no cell --- although
deepseek is the stronger language model, so predictive quality and ratio quality
are distinct (Appendix~\ref{app:ablation-extra}).

\subsection{Is the signal just formatting?}
\label{sec:formatting}

Static formatting statistics separate human and machine code to a degree
\citep{nirob2026whitespaces}, so we bound that explanation from both sides
(Table~\ref{tab:ws}, Appendix~\ref{app:ws}): a whitespace-only logistic probe
trained on the same corpus as the supervised baselines reaches $0.60$ average
AUROC, the static ceiling, while normalizing all formatting away --- provably
leaving the non-whitespace character sequence identical --- costs the detector
$3.5$ AUROC points and still leaves the pure cells $25$--$30$ points above their
per-cell ceilings. The separation between methods in Table~\ref{tab:main} is
therefore not layout.

\subsection{Robustness on real human code}
\label{sec:results:fpr}

A detector is only deployable if its false-positive rate on genuinely
human-written code is low, measured on code that predates widespread model
assistance rather than on the human side of a synthetic pair. Across $19{,}445$
files from repositories created before 2022, all of them negatives, at thresholds
fixed on the development split and not re-tuned, all four read-outs stay between
$0.55$ and $0.81\%$ against a nominal $1\%$ (Table~\ref{tab:fpr},
Appendix~\ref{app:fpr}), so transporting the operating point from paired synthetic
data to real repositories costs essentially nothing. Block- and line-level
false-alarm rates on the same corpus are an order of magnitude higher ($9.90\%$ and
$8.89\%$ for blocks at $K=8$ and $K=16$, $17.33\%$ for lines;
Table~\ref{tab:fprgran}): the file-level
aggregation is what suppresses them, so any deployment that surfaces block- or
line-level highlights has to be calibrated separately.

\subsection{External benchmarks}
\label{sec:results:external}

Table~\ref{tab:external-main} reports AICD-Bench and
DroidCollection under the two protocols of Section~\ref{sec:setup:external}; HCA is
in Table~\ref{tab:external-hca}. Appendix~\ref{app:external} fixes
what each benchmark contributes and states the DroidCollection class mapping. Where
a benchmark is natively multi-class we additionally report the native multi-class
number, so that it is never silently compared against a binary one.

\begin{table}[tbp]
\small
\caption{Transfer (P1) and adaptation (P2) on AICD-Bench and DroidCollection
\citep{droidcollection}. Group~A trains nothing, so it has no P2 column and only
its strongest row (DetectCodeGPT) is shown here; the full zero-shot set is in
Table~\ref{tab:external-zs}. ImBD's P2 uses length-matched pairs within each
benchmark rather than the content-equivalent pairs its SPO stage assumes, since no
external benchmark provides them (Appendix~\ref{app:tables}). Aggregation and
per-row exceptions are in Appendix~\ref{app:tables}.}
\label{tab:external-main}
\begin{center}
\scriptsize
\setlength{\tabcolsep}{1.4pt}
\renewcommand{\arraystretch}{0.78}
\begin{tabular}{llcccccc}
\toprule
& \multirow{2}{*}{Method} & \multicolumn{2}{c}{AICD-Bench T1} & \multicolumn{2}{c}{AICD-Bench T3} & \multicolumn{2}{c}{DroidCollection} \\
\cmidrule(lr){3-4}\cmidrule(lr){5-6}\cmidrule(lr){7-8}
& & P1 & P2 & P1 & P2 & P1 & P2 \\
& & \multicolumn{6}{c}{\scriptsize AUROC / TPR@1\%FPR / macro F1 (\%)} \\
\midrule
A & DetectCodeGPT & 0.703 / 6.5 / 49.7 & -- & 0.666 / 0.2 / 40.2 & -- & 0.669 / 3.0 / 35.1 & -- \\
\midrule
\multirow{6}{*}{B}
& CodeT5+ + MLP & 0.471 / 1.3 / 44.8 & 0.629 / 11.6 / 53.5 & 0.462 / 0.7 / 40.7 & 0.927 / 43.3 / 73.5 & 0.521 / 0.8 / 32.7 & 0.982 / 81.5 / 89.7 \\
& CodeBERT & 0.467 / 1.5 / 44.9 & 0.651 / 17.0 / 58.5 & 0.356 / 0.6 / 40.6 & 0.921 / 49.1 / 76.7 & 0.469 / 1.0 / 47.2 & 0.988 / 87.8 / 94.7 \\
& UniXcoder & 0.489 / 1.9 / 45.3 & 0.618 / 12.6 / 55.0 & 0.374 / 0.7 / 40.7 & 0.929 / 35.8 / 68.9 & 0.499 / 1.5 / 50.8 & 0.991 / 90.1 / 96.2 \\
& \shortstack[l]{\emph{Same backbone}\\\quad + LoRA cls.\ head} & 0.654 / 5.8 / 49.1 & 0.636 / 20.8 / 61.3 & 0.550 / 0.7 / 40.7 & 0.967 / 64.6 / 84.6 & \textbf{0.725} / 5.7 / 37.8 & 0.979 / 81.4 / 89.7 \\
& ImBD & 0.606 / 5.2 / 48.5 & 0.549 / 14.9 / 56.9 & 0.564 / 2.4 / 42.4 & \emph{0.399} / 2.4 / 42.4 & 0.645 / 7.1 / 39.3 & 0.683 / 17.4 / 48.7 \\
& DroidDetect & 0.535 / 1.9 / 45.4 & 0.386 / 0.1 / 43.6 & 0.469 / 0.2 / 40.2 & 0.891 / 42.0 / 72.7 & 0.532 / 1.1 / 51.4 & 0.988 / 86.1 / 96.0 \\
\midrule
\multirow{4}{*}{C}
& Ours: token ratio, file mean & 0.527 / 3.8 / 47.2 & \textbf{0.829} / 11.9 / 54.4 & 0.484 / 2.3 / 42.4 & 0.951 / 29.9 / 65.1 & 0.525 / 3.9 / \textbf{51.8} & 0.988 / 87.0 / 96.5 \\
& \quad + block head, $K=8$ & 0.700 / 9.2 / 52.1 & 0.774 / 19.9 / 60.7 & 0.669 / 4.5 / 44.6 & 0.967 / 68.3 / 86.2 & 0.643 / 4.8 / 35.4 & 0.996 / 93.8 / 97.0 \\
& \quad + block head, $K=16$ & \textbf{0.773} / \textbf{15.9} / \textbf{57.7} & 0.806 / 21.2 / 61.6 & 0.628 / 3.6 / 43.6 & 0.960 / \textbf{73.3} / \textbf{88.5} & 0.603 / 3.3 / 33.3 & 0.996 / 93.5 / 97.0 \\
& \quad + line head, file mean & 0.706 / 15.3 / 57.2 & 0.711 / \textbf{22.6} / \textbf{62.7} & \textbf{0.710} / \textbf{6.4} / \textbf{46.4} & \textbf{0.973} / 73.2 / 88.5 & 0.707 / \textbf{14.5} / 48.7 & \textbf{0.996} / \textbf{94.0} / \textbf{97.0} \\
\bottomrule
\end{tabular}
\end{center}
\end{table}

On DroidCollection the P1 ranking is metric-dependent: the LoRA head has the best
AUROC ($0.725$), our line head the best tail (TPR@1\%FPR $14.5$ against $7.1$ for
ImBD, the strongest baseline tail) and
our token ratio the best macro F1 ($51.8$). Adaptation resolves the ranking --- our
line head leads on all three ($0.996$ / $94.0$ / $97.0$) --- but not the tension
between metrics: the token ratio beats both pretrained encoders on macro F1 once
adapted ($96.5$ vs.\ $94.7$ and $96.2$) while having the weakest tail
($87.0\%$ against $87.8$ and $90.1$), so its errors sit where a low-false-alarm
budget is spent.

The line head's edge under transfer --- the strongest of our read-outs on T3 P1 and
on DroidCollection's P1 --- does not survive adaptation: on T1 P2 it is the weakest
of our own four rows ($0.711$ against $0.829$ for the token ratio), although still
above every Group~B row there. The asymmetry
is expected rather than incidental: T1, T3 and DroidCollection annotate at the file
level, so every line of a positive file carries the same label and the line head has
no localization supervision to exploit --- it degenerates to a file-level classifier
with a mean-pooled loss, and whatever edge it has under transfer comes from the
sequence model rather than from line-level evidence.

On HCA the line head is the strongest of our read-outs at both granularities it
covers: $62.5$ line-level $F1$, and $0.786$ AUROC at the file level, the best
AUROC anywhere in the table ($0.690$ / $0.667$ for the block heads, $0.712$ for
the strongest baseline; Table~\ref{tab:external-hca}). The margin is at the file
level: at the line granularity every method stays near chance, ours at $0.589$.
That fits the segment-length analysis of
Section~\ref{sec:fine}: HCA's median span is two lines and its median AI fraction
is $0.21$, so per-line evidence is thinnest exactly where only the file-level
aggregate carries weight.

\begin{table}[tbp]
\small
\caption{HybridCodeAuthorship \citep{hybridcodeauthorship} under P1, at its native
line granularity, our $K$-line blocks, and the file level. F1 here is
plain, not macro. The weakest zero-shot statistics and the weakest fine-tuned
encoder are omitted here; the full table is
Table~\ref{tab:external-hca-full} in Appendix~\ref{app:external}.}
\label{tab:external-hca}
\begin{center}
\scriptsize
\setlength{\tabcolsep}{3pt}
\renewcommand{\arraystretch}{0.85}
\begin{tabular}{llcccc}
\toprule
& \multirow{2}{*}{Method} & \multicolumn{4}{c}{HCA, P1 (zero adaptation)} \\
\cmidrule(lr){3-6}
& & Line & Block $K{=}8$ & Block $K{=}16$ & File \\
& & \multicolumn{4}{c}{\scriptsize AUROC / TPR@1\%FPR / F1 (\%)} \\
\midrule
\multirow{2}{*}{A}
& Fast-DetectGPT & 0.495 / 1.2 / 2.2 & 0.568 / 1.8 / 3.5 & 0.568 / 1.9 / 3.7 & 0.499 / 0.0 / 0.0 \\
& AIGCode Detector      & 0.646 / 1.4 / 2.7 & 0.513 / 2.2 / 4.1 & 0.501 / 1.7 / 3.2 & 0.496 / 1.4 / 16.7 \\
\midrule
\multirow{5}{*}{B}
& CodeBERT                               & \textbf{0.666} / 6.1 / 11.4 & 0.577 / 4.8 / 9.0 & 0.589 / 5.4 / 10.1 & 0.594 / 3.3 / 6.3 \\
& UniXcoder                              & 0.581 / 3.6 / 6.8 & 0.562 / 4.8 / 9.0 & 0.570 / 5.2 / 9.8 & 0.580 / 2.4 / 4.6 \\
& \emph{Same backbone} + LoRA cls.\ head      & \emph{0.337} / 0.6 / 1.1 & 0.532 / 1.3 / 2.5 & 0.534 / 1.4 / 2.7 & 0.476 / 1.2 / 2.4 \\
& ImBD & 0.536 / 1.3 / 2.5 & 0.531 / 2.6 / 5.0 & 0.553 / 3.0 / 5.7 & 0.472 / 2.2 / 4.3 \\
& DroidDetect & 0.586 / 2.6 / 5.0 & 0.682 / 6.6 / 12.2 & \textbf{0.685} / 9.3 / 16.8 & 0.712 / 11.3 / 26.7 \\
\midrule
\multirow{4}{*}{C}
& Ours: token ratio, line mean           & 0.528 / 0.9 / 53.5 & 0.637 / 6.9 / 49.5 & 0.628 / 9.2 / \textbf{52.1} & 0.638 / 12.8 / 22.7 \\
& \quad + block head, $K=8$              & -- & \textbf{0.685} / \textbf{14.0} / \textbf{51.5} & -- & 0.690 / 15.3 / 26.5 \\
& \quad + block head, $K=16$             & -- & -- & 0.627 / \textbf{11.0} / 50.4 & 0.667 / 14.4 / 25.2 \\
& \quad + line head, run mean            & 0.589 / \textbf{10.6} / \textbf{62.5} & -- & -- & \textbf{0.786} / \textbf{19.1} / \textbf{32.0} \\
\bottomrule
\end{tabular}
\end{center}
\end{table}

\section{Limitations}
\label{sec:limitations}

Our mixed files use few writer models and prompts, and the specification-based
construction strips machine-side dead code and license headers, so the
numbers are an upper bound, not a deployment estimate. Training sees only long AI
segments (median $20$ lines), so the scattered regime is measured but never
trained on, and the positional family, bounded at $-0.001$ AUROC here, is not
ruled out elsewhere. Byte-identical low-entropy pairs put the
same text on both sides as known label noise.

\section{Conclusion}
\label{sec:conclusion}

Detecting AI-involved code is a different problem from detecting AI-written code,
and the difference is not a matter of degree. Where a generated span sits inside
surviving human code, the supervised file-level control that reaches $71.6$--$79.3$
macro F1 on whole machine-written files stays $3.6$ and $1.5$ points above the
all-negative floor, against $11.4$ and $5.5$ for the token-level likelihood ratio
(Table~\ref{tab:main}): a windowed head averages a short inserted span away, while
the ratio between an AI-side and a human-side adapter keeps it addressable and the
line and block heads read it where it sits. The mechanisms are therefore
complementary rather than competing.

What bounds a localizing read-out is the size of the machine-written part relative
to the prediction unit, not the choice of unit: below $\rho = 0.2$ the token ratio
and both block heads lie within $0.8$ macro F1 of each other, and once the median
AI segment falls below any usable block --- two lines on \textsc{refac.} --- the
ordering of $K$ reverses and the line head drops to $54.8$ file-level macro F1 at
$48.0$ line-level F1, so what remains is pooling rather than attribution. The
transfer--adaptation gap carries the same message one level up: most of what
separates the classes on AICD-Bench and DroidCollection is specific to how each
corpus was generated rather than a property of machine-written code that survives
a change of protocol.

\subsubsection*{AI Use Statement}

LLMs were used in this work in the following limited capacities.
\begin{enumerate}
\item \textbf{Dataset construction.} The AI-generated code portion of our internal
dataset was constructed by calling LLM APIs. The training, validation and main test
splits use four writer models --- Claude Opus 5, DeepSeek-V4-Flash, GPT-5.6 and
GLM-5.2; the held-out-model generalization test additionally uses
Qwen3-Coder-Plus, Doubao-Seed-2.1-Pro, Kimi-K2.6 and MiniMax-M2.7.
\item \textbf{Related-work search.} We used LLMs to assist in searching for and
surveying related work.
\item \textbf{Experiment implementation.} We used a coding agent to implement the
experiment scripts according to our pre-specified experimental plan.
\item \textbf{Translation.} The paper was initially drafted in Chinese and we used
LLMs to translate it into English.
\item \textbf{Writing.} All sections of this paper were initially drafted and
completed by the human authors, who established all core ideas, analyses and
conclusions. LLMs were then used to refine and polish these pre-written drafts.
\end{enumerate}
They were not used for proposing or refining hypotheses, for designing or giving
feedback on methodology or experiments, for formulating mathematical claims or
proofs, for cleaning or reformatting data, for qualitative or thematic data
analysis, or for interpreting results; no other task on the required-disclosure
list applies to this work. We have reviewed all AI-assisted work: the experiment
code produced with the coding agent was verified and tested for correctness by one
of the authors. We take responsibility for the final content of this work,
including text, claims and artifacts produced with the aid of generative AI.
\subsubsection*{Ethics Statement}

Detectors of this kind are deployed in assessment and code-provenance settings
where a false positive is costly, which is why every headline number in this paper
is reported at a low-false-positive operating point and why we measure the
false-positive rate on genuinely human code from repositories that predate
widespread model assistance rather than on the human side of a synthetic pair. Two
consequences of our measurements bear directly on responsible use. First, the
false-alarm rate is language-dependent, and the worst per-language rate we observe
is an order of magnitude above the aggregate, so a single global threshold
disadvantages authors writing in the less-represented languages. Second, block-
and line-level highlights carry much higher false-alarm rates than the file-level
verdict, so surfacing them to a human reviewer without separate calibration would
present unreliable evidence as if it were localized proof. We release no model
that scores individual authors, and the training corpus is built from public
repositories under their original licenses.

\appendix
\section{Appendix}

Table~\ref{tab:appguide} maps each part of this appendix to the main-text section
it resolves. Two parts are prerequisites for the rest rather than extensions of a
single result: Appendix~\ref{app:impl} fixes the configuration behind every number
we report, and Appendix~\ref{app:tables} fixes the conventions every table
inherits together with its per-table exceptions. Appendix~\ref{app:ablation-extra}
collects ablations and one negative result that no main-text claim rests on.
Everything else supplies the per-cell, per-bucket or per-language resolution of a
result the main text states in aggregate.

\begin{table}[h]
\small
\caption{Guide to the appendix. The last column is the main-text section whose
result the appendix resolves, or which defers its material here.}
\label{tab:appguide}
\centering
\scriptsize
\renewcommand{\arraystretch}{1.15}
\begin{tabular}{@{}p{3.2cm}p{7.1cm}p{2.2cm}@{}}
\toprule
Appendix & Contents & Main text \\
\midrule
\ref{app:related} Extended related work
  & zero-shot likelihood statistics, supervised code detectors, two-model contrasts, spans and boundaries in text, mixed and fine-grained authorship resources
  & \ref{sec:intro} \\
\ref{app:acc} Accuracy and F1 at the transported threshold
  & why neither metric appears in the main text; accuracy per test cell and per synthesis form against the degenerate baseline
  & \ref{sec:setup:metrics} \\
\ref{app:block} Block-level metrics
  & block and line metrics of both heads, per test cell, at both $K$
  & \ref{sec:fine} \\
\ref{app:loc} Localization diagnostics
  & localization value by AI segment length and by AI ratio $\rho$
  & \ref{sec:fine} \\
\ref{app:ablation-extra} Additional ablations
  & feature families per cell, the aggregation sweep, the line-head read-out, the block-supervision corpus, syntax-aligned blocks, the backbone swap
  & \ref{sec:method:block}, \ref{sec:method:agg}, \ref{sec:ablation} \\
\ref{app:ws} Formatting controls
  & whitespace and length ceilings; the two levels of formatting normalization
  & \ref{sec:formatting} \\
\ref{app:fpr} False positives on real repository code
  & false-alarm rates per language, under the deployment and the accuracy-maximising threshold
  & \ref{sec:results:fpr}, Ethics \\
\ref{app:impl} Implementation details
  & dataset splits, backbone and adapter configuration, scoring passes, head inputs and training
  & \ref{sec:method:lr}, \ref{sec:method:block}, \ref{sec:setup:data}, Reprod. \\
\ref{app:tables} Table conventions and protocol details
  & the conventions every table inherits, per-table exceptions, the main comparison with all three metrics per cell
  & \ref{sec:setup:baselines}, \ref{sec:setup:metrics}, \ref{sec:results:main} \\
\ref{app:external} External benchmarks
  & benchmark composition and class mapping, full zero-shot and HybridCodeAuthorship tables
  & \ref{sec:setup:external}, \ref{sec:results:external} \\
\bottomrule
\end{tabular}
\end{table}

\subsection{Extended related work}
\label{app:related}

This section is the related-work discussion deferred from the main text.

\paragraph{Zero-shot likelihood detectors.}
Zero-shot detectors threshold a likelihood statistic computed under a reference
model: perturbation curvature, estimated by mask-filling
\citep{mitchell2023detectgpt} or in analytic conditional-probability form
\citep{bao2024fastdetectgpt}; log-rank \citep{gehrmann2019gltr} and its ratios LRR
and NPR \citep{su2023detectllm}; and cross-perplexity between a performer and an
observer model, which cancels much of the prompt-difficulty confound
\citep{hans2024binoculars}. They need no labels and transfer across generators, and
they all rest on the premise that machine text is more predictable than human text
under a reference model.

\paragraph{Supervised detectors for machine-generated code.}
DetectCodeGPT is the code-side perturbation detector, inserting newlines and
whitespace instead of filling masked natural-language spans, on the observation
that machine code is more regular in exactly those positions
\citep{shi2024detectcodegpt}. Most supervised code detectors instead train a
discriminator on labelled source: CodeBERT- and CodeGPT-based classifiers
\citep{gptsniffer,codegptsensor}, detectors built on surface-form, whitespace or
embedding features \citep{nirob2026whitespaces}, and function-level
detectors for Python \citep{idialu2024whodunit} and C programs
\citep{bukhari2023}; all of them assume a labelled pool in the language and style
of the code under test. ImBD aligns a proxy model to machine stylistic preference
by direct preference optimization and then reads a curvature statistic under the
aligned model \citep{song2025imbd}, the strongest recent supervised-scoring
competitor for machine-revised text. Our own supervised controls are described in
Section~\ref{sec:setup:baselines}.

\paragraph{Contrasting two models.}
A difference of two models' log-probabilities over a shared backbone is the score
form used in decoding-time control --- contrastive decoding
\citep{li2023contrastivedecoding}, DExperts \citep{liu2021dexperts} and proxy
tuning \citep{liu2024proxytuning} --- where the two sides are experts steering
generation rather than detectors. What we take from that line is the form of the
score; what we add is its role as a per-token detection statistic and the sequence
heads that read it.

\paragraph{Spans and boundaries in text.}
Below the document level, the nearest formulations come from machine-generated
text detection. SeqXGPT labels machine-written sentences with a token-level
classifier \citep{seqxgpt}; SemEval-2024 Task 8 Subtask C asks where the boundary
between human and machine spans falls inside a mixed document
\citep{semeval2024task8}; M4GT-Bench organizes datasets and domains around the
same mixed-text setting \citep{wang2024m4gtbench}, and RoFT studies where human
readers place such boundaries \citep{dugan2020roft}. The task is the same here
and the structural assumptions are not. Mixed text is generated forward from one
prompt, so authorship changes at a point in the sequence and the natural
procedures are change-point detection or token tagging. A mixed file is assembled
by local edits against a surrounding context, with fill-in-the-middle insertions
and whole-block rewrites, so AI involvement is block-aligned rather than a single
cut. That difference is what our block and line heads encode, and it is why we
report attribution at those two granularities instead of a boundary position.

\paragraph{Mixed and fine-grained authorship.}
Most resources for code authorship operate at the file level: DroidCollection is a
multi-generator, multi-domain corpus annotated over four authorship classes, which
also ships the fine-tuned encoder detector DroidDetect \citep{droidcollection}, and
AICD-Bench stays at the file level and targets out-of-distribution robustness
through its own tracks and splits \citep{aicdbench}. HybridCodeAuthorship is the
exception: it provides per-line labels for human--AI co-written files with no
training split \citep{hybridcodeauthorship}, with a median AI segment of two lines.

\subsection{Accuracy and F1 at the transported threshold}
\label{app:acc}

Neither accuracy nor F1 appears in the main text, because the threshold both
require does not transport. The development split contains $8.9\%$ mixed files
and the test set $32.8\%$, so a threshold that maximizes accuracy on development
sits in the wrong place on test, and the test set is additionally imbalanced at
$4.63{:}1$ positives to negatives. Under those conditions the degenerate
all-positive classifier reaches $0.903$ F1 and $0.822$ accuracy on the internal
test set, and all three of our read-outs fall below both: $0.795$, $0.810$ and
$0.797$ F1, and $0.717$, $0.734$ and $0.719$ accuracy, for the token ratio and
the two block heads. This is the reason the main comparison uses AUROC and TPR at
a fixed FPR: both are computable without a transported threshold, and neither can
be won by a degenerate predictor.

Table~\ref{tab:acccell} resolves that same threshold per cell and per synthesis
form, and it has to be read across cells rather than down a column. Every cell
except \textsc{human} contains positives only, so accuracy there is recall and the
all-positive baseline is $1.000$ by construction; the \textsc{human} column is
$1$ minus the false-alarm rate at the same threshold and is the only column that
constrains the threshold from the other side. The per-form panel is where the
degenerate baseline carries information, because its \emph{pure} column pools the
pure-AI files with the human pool ($0.736$ all-positive accuracy): the token ratio
clears that floor at $0.910$, while all three mixed forms sit far below the
$1.000$ their own column demands --- $0.127$ for the token ratio on
fill-in-the-middle, against $0.275$ for the $K{=}8$ head, which is the same
ordering the AUROCs of Table~\ref{tab:main} give and the reason a localizing
read-out is worth its cost on the from-scratch cells.

\begin{table}[tbp]
\small
\caption{Accuracy at the transported accuracy-maximising threshold, per test cell
(upper panel) and per synthesis form (lower panel). Each read-out uses its own
threshold, frozen on the development split by maximizing development accuracy and
applied unchanged to test; the three values are those of the lower panel of
Table~\ref{tab:fpr}. \emph{all-pos.}\ is the degenerate all-positive accuracy,
i.e.\ the floor of that column. Every cell except \textsc{human} holds positives
only, so its accuracy equals recall and its floor is $1.000$; the \textsc{human}
column is $1$ minus the false-alarm rate. In the lower panel \emph{pure} is the
$9{,}619$ files with no mixing --- the three from-scratch cells plus the human
pool --- which is why its floor is $0.736$ rather than $1.000$. The file-level
accuracies quoted in the text are these per-cell values weighted by cell size.}
\label{tab:acccell}
\begin{center}
\scriptsize
\setlength{\tabcolsep}{3pt}
\begin{tabular}{lrccccccc}
\toprule
Read-out & threshold & \textsc{base} & \textsc{unseen} & \textsc{rewrite} & \textsc{compl.} & \textsc{refac.} & \textsc{fim} & \textsc{human} \\
\midrule
Token ratio, file mean       & $0.00081$ & 0.897 & 0.896 & 0.893 & 0.263 & 0.506 & 0.127 & 0.949 \\
\quad + block head, $K{=}8$  & $-1.434$  & 0.844 & 0.911 & 0.804 & 0.490 & 0.527 & 0.275 & 0.923 \\
\quad + block head, $K{=}16$ & $-2.107$  & 0.846 & 0.909 & 0.803 & 0.422 & 0.487 & 0.212 & 0.943 \\
\midrule
\emph{all-pos.}\ floor       & --        & 1.000 & 1.000 & 1.000 & 1.000 & 1.000 & 1.000 & 0.000 \\
\bottomrule
\end{tabular}

\vspace{4pt}
\scriptsize
\setlength{\tabcolsep}{5pt}
\begin{tabular}{lcccc}
\toprule
& \emph{pure} & continuation & refactor diff & fill-in-middle \\
Read-out & ($n = 9{,}619$) & ($1{,}496$) & ($1{,}899$) & ($1{,}300$) \\
\midrule
Token ratio, file mean       & 0.910 & 0.263 & 0.506 & 0.127 \\
\quad + block head, $K{=}8$  & 0.874 & 0.490 & 0.527 & 0.275 \\
\quad + block head, $K{=}16$ & 0.879 & 0.422 & 0.487 & 0.212 \\
\midrule
\emph{all-pos.}\ floor       & 0.736 & 1.000 & 1.000 & 1.000 \\
\bottomrule
\end{tabular}
\end{center}
\end{table}

\subsection{Block-level metrics}
\label{app:block}

Block-level AUROC and F1 for each $K$. All block labels follow
Section~\ref{sec:problem}: a block is positive if it contains at least one AI
line. On the internal test set this makes $72.0\%$ of blocks positive at $K=8$ and
$74.6\%$ at $K=16$, and the block head reaches $0.874$ AUROC at $K=8$ and $0.882$
at $K=16$ over all blocks, against $0.906$ and $0.930$ when the metric is
restricted to blocks that are pure in content. At the development-frozen
threshold its F1 is $82.8$ ($K=8$) and $80.4$ ($K=16$), which is below the
all-positive floors of $83.7$ and $85.5$: at a positive rate above $0.7$, F1
rewards the degenerate classifier, and AUROC is the comparable reading at this
granularity.

Table~\ref{tab:blockcell} resolves both $K$ by cell, and it separates two things
the pooled numbers above conflate. Within a cell the smaller block is at least as
good on all three mixed cells ($0.912$, $0.648$ and $0.650$ at $K=8$ against
$0.897$, $0.646$ and $0.594$ at $K=16$), which is the ordering the lower panel of
Table~\ref{tab:seglen} also shows; pooled over all blocks the ordering reverses
($0.874$ against $0.882$), because pooling mixes blocks from different cells and
the easy separation between the all-positive blocks of the pure cells and the
all-negative blocks of the \textsc{human} cell dominates the ranking. Block size
therefore has to be chosen inside a cell, not on the pooled number.

\begin{table}[tbp]
\small
\caption{Block-level metrics of the block head per test cell, at both $K$. Block
labels are those of Section~\ref{sec:problem}: positive if the block contains at
least one AI line. \emph{pos} is the block
positive rate; AUROC ranks blocks inside the cell and is undefined where every
block carries the same label; $F1$ and \emph{recall} use the block-level threshold
frozen on the development split ($0.3214$ at $K=8$, $0.2271$ at $K=16$) and
transported unchanged. The
\textsc{human} cell has no positive block, so its $F1$ and recall are
undefined; it fixes the negative pool of the pooled numbers quoted above.}
\label{tab:blockcell}
\begin{center}
\scriptsize
\setlength{\tabcolsep}{4pt}
\begin{tabular}{lrcccc}
\toprule
Cell & blocks & pos & AUROC & $F1$ & recall \\
\midrule
\multicolumn{6}{l}{\emph{$K = 8$}, development threshold $0.3214$} \\
\textsc{base}    & 31{,}369 & 1.000 & --    & 89.6 & 81.2 \\
\textsc{unseen}  & 35{,}225 & 1.000 & --    & 92.9 & 86.7 \\
\textsc{rewrite} & 23{,}777 & 1.000 & --    & 86.6 & 76.4 \\
\textsc{compl.}  & 19{,}131 & 0.570 & 0.912 & 80.0 & 71.3 \\
\textsc{refac.}  & 22{,}010 & 0.655 & 0.648 & 62.6 & 53.0 \\
\textsc{fim}     & 14{,}379 & 0.639 & 0.650 & 45.8 & 31.9 \\
\textsc{human}   & 27{,}607 & 0.000 & --    & --   & --   \\
\midrule
\multicolumn{6}{l}{\emph{$K = 16$}, development threshold $0.2271$} \\
\textsc{base}    & 16{,}328 & 1.000 & --    & 89.2 & 80.5 \\
\textsc{unseen}  & 18{,}249 & 1.000 & --    & 92.7 & 86.4 \\
\textsc{rewrite} & 12{,}388 & 1.000 & --    & 84.5 & 73.2 \\
\textsc{compl.}  & 9{,}952  & 0.628 & 0.897 & 71.0 & 56.8 \\
\textsc{refac.}  & 11{,}483 & 0.762 & 0.646 & 58.2 & 44.0 \\
\textsc{fim}     & 7{,}516  & 0.728 & 0.594 & 34.5 & 21.5 \\
\textsc{human}   & 14{,}460 & 0.000 & --    & --   & --   \\
\bottomrule
\end{tabular}
\end{center}
\end{table}

The line head is reported per cell in
Table~\ref{tab:linecell}, since its line-level behaviour is what bounds a
localize-first pipeline. Line-level AUROC is defined only on the three mixed
cells: in the pure cells every line is AI and in the \textsc{human} cell none is.
The pure rows therefore carry a recall but no ranking metric, and the
\textsc{human} row carries neither --- with no positive line its recall and $F1$
are undefined as well, and it appears only to fix the negative pool.

\begin{table}[tbp]
\small
\caption{Line head, per cell. Line labels are those of
Section~\ref{sec:problem}. \emph{Line AUROC} ranks lines
within the cell; it is undefined where all lines carry the same label
(\textsc{base}, \textsc{unseen}, \textsc{rewrite}, \textsc{human}). $F1$ uses the
threshold $-0.272$, selected on the development
splits by line-level $F1$ and transported unchanged. Recall is at the same
threshold. A dash means undefined: the
\textsc{human} cell contains no positive line, so its recall and $F1$
have no value, and the quantity that would take their place --- a
line-level false-alarm rate --- is not reported at this granularity.}
\label{tab:linecell}
\begin{center}
\scriptsize
\setlength{\tabcolsep}{3.5pt}
\begin{tabular}{lrrcccc}
\toprule
& & & AI-line & Line & Line & Line \\
Cell & files & lines & rate & AUROC & $F1$ & recall \\
\midrule
\textsc{base}    & 2{,}541 & 237{,}027 & 1.000 & -- & 91.2 & 83.8 \\
\textsc{unseen}  & 2{,}541 & 267{,}667 & 1.000 & -- & 93.4 & 87.6 \\
\textsc{rewrite} & 1{,}996 & 178{,}961 & 1.000 & -- & 86.2 & 75.7 \\
\textsc{compl.}  & 1{,}496 & 143{,}999 & 0.519 & 0.927 & 85.2 & 90.7 \\
\textsc{refac.}  & 1{,}899 & 165{,}466 & 0.334 & 0.599 & 48.0 & 61.2 \\
\textsc{fim}     & 1{,}300 & 107{,}925 & 0.453 & 0.695 & 58.7 & 55.1 \\
\textsc{human}   & 2{,}541 & 206{,}146 & 0.000 & -- & -- & -- \\
\bottomrule
\end{tabular}
\end{center}
\end{table}

The line-level signal is regime-dependent rather than uniformly weak: $0.927$ on
\textsc{compl.}, where the AI part is one long tail segment, against $0.599$ on
\textsc{refac.}, whose median segment is two lines, and $0.695$ on \textsc{fim}.
On \textsc{refac.}\ the transported line-level $F1$ is $48.0$; on
\textsc{compl.}\ it reaches $85.2$.

\subsection{Localization diagnostics}
\label{app:loc}

Table~\ref{tab:seglen} is the evidence behind Section~\ref{sec:fine}.
Its upper panel separates the aggregation question from
the difficulty question. \emph{Token ratio, file} is the deployed read-out;
\emph{oracle} restricts the positive statistic to the tokens inside the true AI
spans while holding the negative pool fixed, so the step from \emph{file} to
\emph{oracle} measures what the human majority costs the file-level average. It is
a positive-side ablation and not a like-for-like comparison --- a sub-mean over a
ground-truth-selected token subset against a whole-file mean --- so we do not read
it as a performance number. \emph{oracle-pair} additionally replaces the negative
with the same line range of the file's paired human original, which makes both
sides the same statistic and controls for content difficulty, and that is the only
variant we read as an upper bound on localization value: a short-segment bucket is
itself composed of low-entropy files, so \emph{oracle} alone overstates the signal.
The \emph{length-only} probe is reported for the same reason
at the file level. The lower panel reads the block head at its own granularity and
at the file level from the mean of the same logits, so that the two effects of the
head (better attribution and better file-level pooling) can be told apart.

\begin{table}[tbp]
\small
\caption{Localization value and block size by segment length, on the two
interior-segment cells \textsc{compl.} and \textsc{fim} (pooled). A file's bucket
is the median length, in non-blank lines, of its AI segments.
\emph{Upper panel:} the token-level ratio aggregated over the whole file, over
the true AI spans (\emph{oracle}), and against the same line range of the paired
human original (\emph{oracle-pair}, the strictest control and the only variant we
read as an upper bound on localization value); \emph{length-only} is a logistic
probe on line and character count, included as a difficulty control.
\emph{Lower panel:} the block head read at block level, against the block labels
of Section~\ref{sec:problem}, and at file level, by the mean of the same
logits. All entries are threshold-free AUROC. Segment merging
and the negative pools are defined in Appendix~\ref{app:tables}.}
\label{tab:seglen}
\begin{center}
\scriptsize
\setlength{\tabcolsep}{4pt}
\begin{tabular}{lccccc}
\toprule
Variant & $\le 4$ & 5--9 & 10--24 & $\ge 25$ & All \\
\midrule
\#positive files         & 439 & 449 & 756 & 1{,}152 & 2{,}796 \\
\midrule
\multicolumn{6}{l}{\emph{Token-level aggregation}} \\
Token ratio, file        & 0.510 & 0.561 & 0.630 & 0.775 & 0.660 \\
Token ratio, oracle      & 0.650 & 0.738 & 0.840 & 0.920 & 0.827 \\
Token ratio, oracle-pair & 0.560 & 0.654 & 0.771 & 0.888 & 0.758 \\
Length-only probe        & 0.453 & 0.497 & 0.543 & 0.680 & 0.578 \\
\midrule
\multicolumn{6}{l}{\emph{Block head (choosing $K$)}} \\
Block AUROC, $K=8$       & 0.538 & 0.615 & 0.718 & 0.900 & 0.803 \\
Block AUROC, $K=16$      & 0.520 & 0.606 & 0.701 & 0.883 & 0.776 \\
File AUROC, $K=8$        & 0.549 & 0.596 & 0.722 & 0.889 & 0.743 \\
File AUROC, $K=16$       & 0.531 & 0.580 & 0.691 & 0.863 & 0.719 \\
\bottomrule
\end{tabular}
\end{center}
\end{table}

\paragraph{Stratification by AI ratio.} Table~\ref{tab:rho} reads the four
Group~C rows of Table~\ref{tab:main} against $\rho$ rather than against segment
length, on the three mixed cells pooled. All four rise monotonically with $\rho$,
and the unit-level heads separate from the token ratio only above
$\rho \approx 0.4$: in the lowest bucket the four read-outs lie within $0.022$
AUROC of each other ($0.642$--$0.664$), while at $\rho \ge 0.8$ the $K{=}8$ head
is $0.051$ AUROC and $21.0$ TPR points above the token ratio. The stratification
is therefore a second statement of the bound of Section~\ref{sec:fine}: a
supervised unit-level head buys little when the machine-written part of the file
is small, whether ``small'' is measured as a short segment or as a low ratio.

\begin{table}[tbp]
\small
\caption{File-level detection stratified by the AI ratio $\rho$, on the three
mixed-authorship cells \textsc{compl.}, \textsc{refac.}\ and \textsc{fim}
(pooled). A file's column is its $\rho$ (Section~\ref{sec:problem}) and the
boundaries are the five construction buckets of Section~\ref{sec:method:data},
except that the lowest bucket extends down to $0$ and the highest includes
$\rho = 1$: \textsc{refac.}\ is produced by refactoring diffs whose coverage is
not solved backwards from a target $\rho$, so $177$ of its files fall below
$0.05$ and $15$ at or above $0.95$. Each entry is AUROC / TPR@1\%FPR / macro F1
(\%). \tabconv{} Bold marks the highest AUROC of a column. The positive set is
the files of that column; the negative pool is the shared \textsc{human} cell
($\Nhuman$ files) in every column, and the threshold behind both
operating-point metrics is the single $99$th percentile of its scores, so
TPR@1\%FPR is comparable across columns. Chance is AUROC $0.500$ and TPR
$1.0\%$. Macro F1 is \emph{not} comparable across columns, because the positive
count changes while the negative pool does not; its all-negative reference, the
chance level at this operating point, is given as a row.}
\label{tab:rho}
\begin{center}
\scriptsize
\setlength{\tabcolsep}{1.5pt}
\begin{tabular}{lcccccc}
\toprule
Method & $<0.2$ & $0.2$--$0.4$ & $0.4$--$0.6$ & $0.6$--$0.8$ & $\ge 0.8$ & All \\
\midrule
\#positive files       & 1{,}298 & 1{,}105 & 914 & 801 & 577 & 4{,}695 \\
all-negative macro F1  & 39.8 & 41.1 & 42.4 & 43.2 & 44.9 & 26.0 \\
\midrule
Token ratio, file mean & 0.642/11.2/50.4 & 0.702/17.8/57.0 & 0.716/17.6/57.9 & 0.778/23.3/62.8 & 0.858/27.0/66.5 & 0.720/18.0/43.5 \\
\quad + block head, $K{=}8$  & \textbf{0.664}/11.3/50.5 & \textbf{0.731}/18.6/57.6 & 0.796/22.3/61.4 & \textbf{0.858}/35.0/70.4 & \textbf{0.909}/48.0/78.4 & \textbf{0.769}/23.7/48.2 \\
\quad + block head, $K{=}16$ & 0.653/12.1/51.2 & 0.727/19.9/58.6 & 0.786/20.4/60.0 & 0.842/32.0/68.5 & 0.901/43.2/75.9 & 0.759/22.7/47.4 \\
\quad + line head, file mean & 0.646/8.3/47.8 & 0.713/12.6/52.7 & \textbf{0.799}/14.9/55.7 & 0.843/23.1/62.6 & 0.900/38.3/73.3 & 0.756/16.8/42.4 \\
\bottomrule
\end{tabular}
\end{center}
\end{table}

\subsection{Additional ablations}
\label{app:ablation-extra}

These tables extend Table~\ref{tab:ablation} to the components whose ablation does
not change the main conclusions: the per-cell reading of the feature families, the
aggregation sweep, the line-head read-out, the construction of the block
supervision, and one negative result on block partitioning. Unless a caption says
otherwise the protocol is that of Table~\ref{tab:ablation}: file-level scores, the
shared \textsc{human} cell ($\Nhuman$ files) as the negative pool, $\Delta$ as a
paired bootstrap difference against the panel's reference row ($1{,}000$
replicates, $95\%$ interval, threshold recomputed inside each replicate), and
$^{\ast}$ for an interval that excludes zero. Chance is AUROC $0.500$ and TPR
$1.0\%$.

\paragraph{Feature families per cell.} Table~\ref{tab:dropcell} resolves the four
family rows of Table~\ref{tab:ablation} by cell, which is where a shortcut to the
construction prior would have to show up.

\begin{table}[tbp]
\small
\caption{Feature-family ablations of the $K{=}16$ block head, per test cell. The
first row is the full head's AUROC on each cell; the remaining rows are
\emph{paired differences} of a head retrained without that family, computed on
that cell alone ($^{\ast}$: $95\%$ bootstrap interval excludes zero). Each column
scores that cell's positive files against the shared \textsc{human} cell. The prior
that the positional family could
exploit is cell-specific: the AI segment is always at the tail in \textsc{compl.},
always in the interior in \textsc{fim}, and scattered over the whole file in
\textsc{refac.}}
\label{tab:dropcell}
\begin{center}
\scriptsize
\setlength{\tabcolsep}{4pt}
\begin{tabular}{lcccccc}
\toprule
& \textsc{base} & \textsc{unseen} & \textsc{rewrite} & \textsc{compl.} & \textsc{refac.} & \textsc{fim} \\
\midrule
Block head $K{=}16$ (AUROC) & 0.954 & 0.974 & 0.944 & 0.789 & 0.819 & 0.638 \\
\midrule
$\Delta$ no raw family        & $+0.007^{\ast}$ & $-0.007^{\ast}$ & $+0.023^{\ast}$ & $-0.036^{\ast}$ & $-0.004$ & $-0.031^{\ast}$ \\
$\Delta$ no $z$-score family  & $-0.008^{\ast}$ & $-0.010^{\ast}$ & $-0.010^{\ast}$ & $-0.014^{\ast}$ & $-0.026^{\ast}$ & $+0.004$ \\
$\Delta$ no positional family & $+0.001$ & $-0.001$ & $+0.001$ & $-0.005^{\ast}$ & $-0.001$ & $+0.001$ \\
$\Delta$ no language family   & $+0.000$ & $-0.000$ & $+0.004^{\ast}$ & $+0.006^{\ast}$ & $+0.001$ & $+0.007^{\ast}$ \\
\bottomrule
\end{tabular}
\end{center}
\end{table}

Neither the positional nor the language family moves any cell by more than
$0.007$, and on \textsc{compl.}, the cell whose segment position is perfectly
predictable, removing the positional family costs $0.005$. The raw and
$z$-score families are complementary rather than redundant: removing the raw
family \emph{improves} \textsc{base} and \textsc{rewrite} (up to $+0.023$ on
\textsc{rewrite}) and costs $0.007$ on \textsc{unseen}, while costing the most on
the two interior-segment cells
($-0.036$ on \textsc{compl.}, $-0.031$ on \textsc{fim}), so what it contributes is
specific to reading a boundary inside a file; removing the within-file $z$-scores
costs $0.008$--$0.026$ on five of the six cells instead.

\paragraph{Aggregation.} Table~\ref{tab:agg} sweeps the six file-level
aggregations of the block probabilities and the four token-level aggregations of
Section~\ref{sec:method:agg} against one common reference. The ten read-outs are
implemented as follows. Over the per-block probabilities $\hat z_b$ of a file's
block set $\mathcal{B}$: \emph{logit mean} averages $\sigma^{-1}(\hat z_b)$ after
clipping the probabilities away from $0$ and $1$ (Equation~\ref{eq:agg});
\emph{mean} averages $\hat z_b$ in probability space instead; \emph{top-$25\%$}
and \emph{top-$10\%$ mean} sort the blocks by $\hat z_b$ and average the highest
$\lceil 0.25|\mathcal{B}|\rceil$ and $\lceil 0.10|\mathcal{B}|\rceil$ of them, at
least one block in either case; \emph{max} takes the single highest $\hat z_b$,
which carries a length bias the two top-share means do not, since a file with
more blocks has a systematically higher extreme; and \emph{best adjacent pair} is
$\max_b \min(\hat z_b, \hat z_{b+1})$, a ``two blocks in a row are high''
statistic that encodes the prior that an AI segment covers more than one block
and is less sensitive to a single noisy block than the max. Over the per-token
ratio $s_t$: \emph{file mean} is Equation~\ref{eq:agg}; \emph{sliding-window max}
takes the largest within-window mean of $s_t$ over $128$-token windows at stride
$32$, falling back to the file mean for files shorter than one window;
\emph{change-point tail} places a single cut with a CUSUM statistic on $s_t$ and
averages the tokens after it, keeping the file mean when fewer than $8$ tokens
remain; and \emph{best segment} splits the file by recursive binary segmentation
--- at most $8$ cuts, each segment at least $16$ tokens, and a split accepted
only if its CUSUM statistic clears a gate fixed on the development split ---
and takes the largest per-segment mean.

\begin{table}[tbp]
\small
\caption{Aggregation sweep. Upper block: the $K{=}16$ block head read out by the
six aggregations of its per-block probabilities defined above. Lower block: the
token-level ratio under the four aggregations of
Section~\ref{sec:method:agg}. All $\Delta$ are paired differences
against the single reference row, so the lower block's $\Delta$ contains both the
change of read-out and the change of aggregation. Columns, protocol and
$^{\ast}$ are as in Table~\ref{tab:ablation}.}
\label{tab:agg}
\begin{center}
\scriptsize
\setlength{\tabcolsep}{1pt}
\begin{tabular}{lcccccc}
\toprule
& \multicolumn{2}{c}{All cells} & \multicolumn{2}{c}{Mixed cells} & \multicolumn{2}{c}{TPR@1\%FPR} \\
\cmidrule(lr){2-3}\cmidrule(lr){4-5}\cmidrule(lr){6-7}
Aggregation & AUROC & $\Delta$ & AUROC & $\Delta$ & \% & $\Delta$ \\
\midrule
\multicolumn{7}{l}{\emph{Block head}, $K{=}16$} \\
logit mean (ref.)        & 0.853 & -- & 0.749 & -- & 48.4 & -- \\
\quad mean               & 0.853 & \ci{-0.001}{-0.002}{+0.001} & 0.752 & \cis{+0.004}{+0.001}{+0.006} & 47.9 & \ci{-0.5}{-2.2}{+1.8} \\
\quad top-$25\%$ mean    & 0.850 & \cis{-0.003}{-0.005}{-0.000} & 0.754 & \cis{+0.006}{+0.001}{+0.010} & 45.2 & \ci{-3.2}{-5.3}{+0.1} \\
\quad top-$10\%$ mean    & 0.849 & \cis{-0.004}{-0.007}{-0.001} & 0.753 & \ci{+0.005}{-0.000}{+0.010} & 44.2 & \cis{-4.2}{-5.9}{-0.2} \\
\quad max                & 0.848 & \cis{-0.005}{-0.008}{-0.002} & 0.752 & \ci{+0.004}{-0.001}{+0.009} & 44.1 & \cis{-4.3}{-6.1}{-0.3} \\
\quad best adjacent pair & 0.846 & \cis{-0.007}{-0.010}{-0.003} & 0.754 & \ci{+0.005}{-0.001}{+0.012} & 42.8 & \cis{-5.6}{-7.5}{-1.8} \\
\midrule
\multicolumn{7}{l}{\emph{Token ratio} $\slr$} \\
file mean                & 0.838 & \cis{-0.015}{-0.021}{-0.010} & 0.707 & \cis{-0.041}{-0.051}{-0.032} & 47.1 & \ci{-1.3}{-4.9}{+3.2} \\
\quad sliding-window max & 0.827 & \cis{-0.026}{-0.034}{-0.018} & 0.715 & \cis{-0.034}{-0.046}{-0.021} & 41.7 & \cis{-6.7}{-10.2}{-2.0} \\
\quad change-point tail  & 0.820 & \cis{-0.033}{-0.040}{-0.025} & 0.704 & \cis{-0.044}{-0.055}{-0.033} & 22.4 & \cis{-26.0}{-31.0}{-16.5} \\
\quad best segment       & 0.838 & \cis{-0.016}{-0.021}{-0.010} & 0.707 & \cis{-0.041}{-0.051}{-0.031} & 47.0 & \ci{-1.4}{-4.8}{+3.4} \\
\bottomrule
\end{tabular}
\end{center}
\end{table}

At the block level the entire choice is worth $0.007$ AUROC, and the
development-selected logit mean is not the optimum on the mixed cells, where
plain mean, top-$25\%$ and $K{=}8$ are $0.004$--$0.013$ higher; we keep the
development choice rather than reporting the test-optimal one. At the token level
the choice matters at the operating point rather than in AUROC: the change-point
tail mean keeps $0.820$ AUROC but loses more than half of the TPR at $1\%$ FPR
($47.1 \to 22.4$), whereas the best-segment variant is indistinguishable from the
plain mean. Neither of the two localization-flavoured aggregations of the token
score buys anything over averaging it, which is the reason the block head, and
not a segmentation of the token sequence, is our attribution read-out.

\paragraph{Line-head read-out and its positional information.}
The line head consumes the same token profile as the block head and differs only
in supervision granularity, so its read-out is ablated separately
(Table~\ref{tab:linehead}). The learned aggregator over line states is a small
Transformer, which is equivariant to a permutation of the lines: the position of a
line inside the file can only reach it through the sinusoidal encoding, and
removing that encoding is therefore the positional ablation on this path.

\begin{table}[tbp]
\small
\caption{Line-head read-out. Panel~1: aggregations of the per-line logits of the
line head reported in Table~\ref{tab:main}, with the $K{=}16$ block head added as
the other path over the same token profile. Panels~2 and~3: the learned attention
aggregator, with and without the sinusoidal positional encoding. Each panel is
compared against \emph{its own} line-mean row, because the line head has to be
retrained for each of the three runs and cuDNN GRU kernels are not bitwise
reproducible: per-file scores of two retrainings correlate at $0.99$ but are not
identical, so absolute values are comparable within a panel and not across
panels. Columns, protocol and $^{\ast}$ are as in Table~\ref{tab:ablation}.}
\label{tab:linehead}
\begin{center}
\scriptsize
\setlength{\tabcolsep}{1pt}
\begin{tabular}{lcccccc}
\toprule
& \multicolumn{2}{c}{All cells} & \multicolumn{2}{c}{Mixed cells} & \multicolumn{2}{c}{TPR@1\%FPR} \\
\cmidrule(lr){2-3}\cmidrule(lr){4-5}\cmidrule(lr){6-7}
Read-out & AUROC & $\Delta$ & AUROC & $\Delta$ & \% & $\Delta$ \\
\midrule
\multicolumn{7}{l}{\emph{Panel 1}: line head of Table~\ref{tab:main}} \\
line mean (ref.)       & 0.836 & -- & 0.751 & -- & 37.4 & -- \\
\quad line max         & 0.832 & \cis{-0.004}{-0.007}{-0.001} & 0.756 & \ci{+0.005}{-0.000}{+0.010} & 37.4 & \ci{+0.0}{-2.8}{+3.5} \\
\quad block head $K{=}16$ & 0.853 & \cis{+0.017}{+0.011}{+0.024} & 0.749 & \ci{-0.003}{-0.013}{+0.007} & 48.4 & \cis{+11.0}{+6.3}{+14.2} \\
\midrule
\multicolumn{7}{l}{\emph{Panel 2}: retrained line head, positional encoding on} \\
line mean (ref.)       & 0.833 & -- & 0.750 & -- & 35.8 & -- \\
\quad attention aggregator & 0.823 & \cis{-0.009}{-0.013}{-0.006} & 0.752 & \ci{+0.002}{-0.002}{+0.006} & 24.2 & \cis{-11.6}{-15.4}{-7.1} \\
\midrule
\multicolumn{7}{l}{\emph{Panel 3}: retrained line head, positional encoding off} \\
line mean (ref.)       & 0.832 & -- & 0.749 & -- & 37.9 & -- \\
\quad attention aggregator & 0.832 & \ci{-0.001}{-0.003}{+0.001} & 0.757 & \cis{+0.007}{+0.005}{+0.010} & 35.2 & \ci{-2.7}{-18.7}{+1.3} \\
\bottomrule
\end{tabular}
\end{center}
\end{table}

Averaging the line logits is the best read-out of the line head: the max is
$0.004$ below it and the learned aggregator $0.009$ below, the latter losing a
third of the TPR at $1\%$ FPR. Removing the positional encoding does not hurt the
aggregator --- it closes the gap to the line mean ($-0.001$, interval containing
zero) and gains $0.007$ on the mixed cells --- so on this path too the model is
not using where in the file a segment sits. Panel~1 also quantifies the
difference between the two supervision granularities at equal input: the block
head is $0.017$ AUROC and $11.0$ TPR points above the line head averaged over all
cells, while on the mixed cells the two are indistinguishable ($-0.003$, interval
containing zero), which is the same trade-off visible per cell in
Table~\ref{tab:main}.

\paragraph{How the block supervision is constructed.}
Section~\ref{sec:method:data} makes two choices that can be ablated by retraining
the head on a different corpus: mixing two synthesis forms, and keeping the paired
human original of every mixed file as a negative. Table~\ref{tab:supervision}
does that, and its own training diagnostics are the reason we read it as a
data-volume result rather than as a construction result.

\begin{table}[tbp]
\small
\caption{Supervision-construction ablation for the $K{=}16$ block head. Every row
retrains the head with the same features and hyper-parameters on a different
training corpus, for a fixed $4$ epochs (the reference's early-stopping point), so
that the rows differ in data and not in schedule; the development split is
unchanged and is used only to pick the aggregation. \emph{Pure blocks} counts the
blocks that are entirely AI or entirely human, i.e.\ those that enter the loss,
and \emph{dev} is the block-level AUROC on the development split. \emph{Corpus
$63\%$} is a hash-selected subsample chosen to match the block count of
\emph{continuation only}, which is the only pair in this table that separates the
synthesis form from the corpus volume. Test columns, protocol and $^{\ast}$
follow Table~\ref{tab:ablation}.}
\label{tab:supervision}
\begin{center}
\scriptsize
\setlength{\tabcolsep}{2.5pt}
\begin{tabular}{lrrccccc}
\toprule
& \multicolumn{3}{c}{Training corpus} & \multicolumn{2}{c}{All cells} & Mixed & TPR@1\% \\
\cmidrule(lr){2-4}\cmidrule(lr){5-6}\cmidrule(lr){7-7}\cmidrule(lr){8-8}
Corpus & files & pure blocks & dev & AUROC & $\Delta$ & AUROC & \% \\
\midrule
full corpus (ref.)        & 3{,}164 & 17{,}767 & 0.914 & 0.853 & -- & 0.749 & 48.4 \\
\quad continuation only   & 1{,}686 & 11{,}111 & 0.818 & 0.730 & \cis{-0.123}{-0.133}{-0.114} & 0.620 & 8.8 \\
\quad fill-in-middle only & 1{,}478 & 6{,}656  & 0.620 & 0.583 & \cis{-0.271}{-0.284}{-0.259} & 0.492 & 1.6 \\
\quad corpus halved       & 1{,}582 & 8{,}637  & 0.754 & 0.686 & \cis{-0.168}{-0.178}{-0.157} & 0.556 & 10.1 \\
\quad corpus $63\%$       & 1{,}994 & 10{,}807 & 0.768 & 0.683 & \cis{-0.170}{-0.181}{-0.161} & 0.558 & 8.0 \\
\quad no paired originals & 1{,}582 & 7{,}220  & 0.716 & 0.777 & \cis{-0.077}{-0.085}{-0.070} & 0.669 & 24.8 \\
\bottomrule
\end{tabular}
\end{center}
\end{table}

The diagnostics have to be read before the $\Delta$ column. The reference stops
improving on development after $4$ epochs at a block-level AUROC of $0.914$;
\emph{every} reduced-corpus run peaks after a single epoch at $0.62$--$0.82$ and
then descends monotonically while its training loss falls to $0.003$, and pinning
all runs to $4$ epochs reproduces the same development AUROC to three decimals.
The head therefore memorizes a reduced corpus, which means these differences
conflate the construction with an optimisation failure, and we report them as a
sensitivity to corpus volume rather than as the effect of the construction.

Within that limit one comparison is still clean. \emph{Continuation only}
($11{,}111$ pure blocks) is \emph{above} the volume-matched subsample that keeps
both forms ($10{,}807$ blocks): $0.730$ against $0.683$. We therefore have no
evidence that mixing continuation with fill-in-the-middle supervision is what the
block head needs; what the numbers support is the weaker statement that it needs
volume ($17{,}767$ blocks $\to 0.853$, roughly $11{,}000$ blocks $\to 0.68$--$0.73$
regardless of form). Removing the paired human originals costs $0.077$ AUROC and
half of the TPR, in the direction Section~\ref{sec:method:data} predicts, but that
corpus is also the smallest of the six, so that row does not separate the paired
negative from the volume either. One further choice is not load-bearing: weighting
the loss so that every file contributes equally, rather than every block, changes
the average AUROC by $+0.001$ and the TPR by $+0.0$ points.

\paragraph{Syntax-aligned blocks: a negative result.}
Fixed $K$-line blocks cut across syntax, so the natural alternative is to make one
block per declaration. We built that partition with \texttt{tree-sitter} --- one
block per terminal declaration, container declarations expanded into their members,
and the lines between declarations merged into \emph{gap} blocks --- and measured
alignment directly, without training a head, on the three mixed cells
(Table~\ref{tab:ast}). It does not align better than the fixed grid.

\begin{table}[tbp]
\small
\caption{Alignment between block boundaries and true AI segments on the three
mixed cells, for four partitions of the same files. The first three columns are
shares of \emph{blocks}: a block is pure-AI or pure-human if all of its lines have
that label, mixed otherwise. \emph{AI lines captured} is the median over files with
at least one AI line of the fraction of that file's AI lines that lie inside
all-AI blocks --- the part of the signal a block-level head can see undiluted ---
and \emph{diluted} is the share of AI lines that sit in mixed blocks, pooled over
the whole cell and weighted by AI lines. The two columns are computed on different
bases (a median over files against a line-weighted pool over the cell), so they
are not complements of each other and do not sum to $100$. A captured median of
$0.0$ means that in more than half of that cell's files no block is entirely AI. \emph{AST} is the declaration partition, \emph{AST+K8 gaps}
re-cuts its gap blocks on the $8$-line grid. Files that \texttt{tree-sitter}
cannot parse ($1.5$--$2.1\%$ per cell) or that contain no declaration
($31$--$36\%$) fall back to one block per file and are counted as they are. No
head is trained here: these are properties of the partition.}
\label{tab:ast}
\begin{center}
\scriptsize
\setlength{\tabcolsep}{4pt}
\begin{tabular}{lccccc}
\toprule
Partition & pure-AI \% & pure-human \% & mixed \% & AI lines captured \% & diluted \% \\
\midrule
\multicolumn{6}{l}{\textsc{compl.}\ ($n = 1{,}496$)} \\
$K = 8$       & 50.1 & 43.0 & 6.9  & 92.3 & 6.5 \\
$K = 16$      & 48.5 & 37.2 & 14.3 & 82.4 & 13.4 \\
AST           & 51.8 & 36.6 & 11.6 & 11.8 & 50.8 \\
AST+K8 gaps   & 49.7 & 44.1 & 6.2  & 90.5 & 18.3 \\
\midrule
\multicolumn{6}{l}{\textsc{refac.}\ ($n = 1{,}899$)} \\
$K = 8$       & 8.8  & 34.5 & 56.7 & 0.0  & 72.9 \\
$K = 16$      & 5.3  & 23.8 & 70.9 & 0.0  & 83.8 \\
AST           & 6.2  & 48.6 & 45.2 & 0.0  & 91.4 \\
AST+K8 gaps   & 9.2  & 46.2 & 44.6 & 0.0  & 77.9 \\
\midrule
\multicolumn{6}{l}{\textsc{fim}\ ($n = 1{,}300$)} \\
$K = 8$       & 24.0 & 36.1 & 39.9 & 28.1 & 44.5 \\
$K = 16$      & 13.7 & 27.2 & 59.1 & 0.0  & 66.9 \\
AST           & 27.1 & 38.8 & 34.1 & 0.0  & 81.5 \\
AST+K8 gaps   & 27.4 & 40.8 & 31.9 & 10.5 & 57.0 \\
\bottomrule
\end{tabular}
\end{center}
\end{table}

The pure declaration partition is unusable: on \textsc{compl.}\ it captures
$11.8\%$ of the AI lines in all-AI blocks against $92.3\%$ for $K=8$, because
$55$--$59\%$ of all lines fall between declarations and are swallowed by long gap
blocks. Re-cutting the gaps on the $8$-line grid fixes that ($90.5\%$) and does
lower the mixed-block share on the two cells where it is high ($44.6$ against
$56.7$ on \textsc{refac.}, $31.9$ against $39.9$ on \textsc{fim}), but it never
captures more AI lines than the plain grid on any of the three cells. The reason is
that the AI segments of this corpus are not declaration-shaped: of the $688$ AI
segments in the Python files of \textsc{fim}, none coincides with a declaration,
$345$ lie strictly inside one, and $343$ cross a declaration boundary or fall
outside every declaration --- the hole positions are drawn to hit a target AI ratio,
and the refilled text shifts every line number after it. We therefore keep the
fixed grid. This is a statement about this benchmark, not about the idea:
a corpus of whole-function completions would align with declarations by
construction and is not measurable here.

\paragraph{Backbone swap.} The deepseek-coder-1.3B row of
Table~\ref{tab:ablation} uses the same recipe, training files and hyperparameters
as the Qwen2.5-Coder-1.5B reference, and the $\Delta$ is significant under the
paired bootstrap. Its tokenizer cuts the same files into $27\%$ more tokens, so
the row is restricted to the token-ratio file-mean read-out; its base NLL on this
corpus is also visibly lower than Qwen's.

\subsection{Formatting controls}
\label{app:ws}

Table~\ref{tab:ws} is the evidence behind Section~\ref{sec:formatting}, and it
bounds the formatting explanation from two sides. The left panel is a static
ceiling: a logistic probe trained on the same corpus as the supervised baselines,
restricted to whitespace features, reaches only $0.64$--$0.67$ AUROC on the pure
cells, $0.54$ on \textsc{compl.}, and chance level on \textsc{fim}; with file
length alone the probe is near chance on every cell.

The right panel intervenes on the input instead: we erase formatting from the
files and re-score them with the unchanged detector. Two levels of normalization
are applied, in the following order.

\emph{Soft} unifies line endings (\texttt{CRLF} and lone \texttt{CR} become
\texttt{LF}), then strips trailing spaces and tabs from every line, then collapses
each run of blank lines to a single blank line and drops the file's leading and
trailing blank lines. Indentation and intra-line whitespace are left untouched, so
the code remains syntactically valid in any language.

\emph{Hard} applies \emph{Soft} first and then, line by line, removes all leading
spaces and tabs, collapses every remaining run of spaces and tabs inside the line
to a single space, and deletes blank lines entirely. This destroys every
whitespace statistic --- indentation width, blank-line structure and alignment
alike (Python files lose their block structure; the scorer reads the file as text
and does not require compilable code).

Every operation at both levels only deletes or merges space, tab and newline
characters and never touches a non-whitespace one, which is why both levels
provably leave the non-whitespace character sequence identical to the original
file. Erasing formatting
entirely leaves the detector far above the whitespace ceiling of the left panel
and above every zero-shot method, at the cost read off in
Section~\ref{sec:formatting}; the one cell that moves the other way is
\textsc{fim}, which \emph{improves} by $1.7$ points, indicating that indentation
was acting as noise there. The detector reads the code, not the layout.

\begin{table}[tbp]
\small
\caption{Is the signal formatting? \emph{Left panel:} logistic probes trained on
the same corpus as the supervised baselines, restricted to nine whitespace
features or to length alone --- the static ceiling. \emph{Right panel:} the
file-mean token-ratio detector of Table~\ref{tab:main}, adapters and aggregation
unchanged, re-scoring the test files after the two levels of formatting
normalization defined above; both leave the non-whitespace character
sequence identical, so any change is attributable to the erased formatting.
$\Delta$ is \emph{Hard} minus \emph{Original}, computed from the unrounded
AUROCs, so it can differ by $0.001$ from the difference of the two rounded
entries. All entries are threshold-free
file-level AUROC against the shared \textsc{human} cell, with no threshold chosen
anywhere. The feature list is in Appendix~\ref{app:tables}.}
\label{tab:ws}
\centering
\scriptsize
\setlength{\tabcolsep}{4.5pt}
\begin{tabular}{lcccccc}
\toprule
& \multicolumn{2}{c}{Static probe (ceiling)} & \multicolumn{4}{c}{Detector, formatting erased} \\
\cmidrule(lr){2-3}\cmidrule(lr){4-7}
Cell & Whitespace only & Length only & Original & Soft & Hard & $\Delta$ (Hard$-$Orig.) \\
\midrule
\textsc{base}    & 0.650 & 0.553 & 0.964 & 0.958 & 0.925 & $-0.039$ \\
\textsc{unseen}  & 0.643 & 0.561 & 0.968 & 0.962 & 0.941 & $-0.028$ \\
\textsc{rewrite} & 0.666 & 0.564 & 0.972 & 0.963 & 0.914 & $-0.058$ \\
\textsc{compl.}  & 0.539 & 0.603 & 0.711 & 0.699 & 0.675 & $-0.036$ \\
\textsc{refac.}  & 0.600 & 0.539 & 0.810 & 0.789 & 0.741 & $-0.069$ \\
\textsc{fim}     & 0.504 & 0.550 & 0.601 & 0.600 & 0.619 & $+0.017$ \\
\midrule
Avg.             & 0.600 & 0.562 & 0.838 & 0.828 & 0.802 & $-0.035$ \\
\bottomrule
\end{tabular}
\end{table}

\subsection{False positives on real repository code}
\label{app:fpr}

Table~\ref{tab:fpr} is the evidence behind Section~\ref{sec:results:fpr}, and
Table~\ref{tab:fprgran} carries the block- and line-granularity rates quoted at
the end of that section. Two
threshold regimes are reported because they answer different questions: the $1\%$
FPR operating point is the deployment setting used throughout the paper, and the
accuracy-maximising threshold is the one behind Table~\ref{tab:acccell}
(Appendix~\ref{app:acc}). The column ``synthetic'' repeats the same computation on
the $\Nhuman$ human files of the internal test set, so the two columns isolate the
effect of changing the negative pool from synthetic to real repository code.

At the deployment operating point all four read-outs stay at or below $0.81\%$,
against a nominal $1\%$ on the development split and $0.35$--$0.87\%$ on the
synthetic negatives, and the block head at $K=16$ is the most conservative
read-out on both pools. The accuracy-maximising threshold sits in the middle of
the score distribution by construction and therefore gives $4.7$--$10.0\%$; that
regime is reported for completeness, not as an operating point. Per-language rates
are read off the token ratio, the only read-out whose ordering is stable:
JavaScript ($0.86\%$), PHP ($1.42\%$) and Ruby ($1.23\%$) at the loose end and Go
($0.16\%$), Java ($0.28\%$) and Python ($0.33\%$) at the tight end, the same
ordering as on the synthetic pool. The unit-level heads do not preserve it --- Go
is the loosest of the five tabulated languages under the $K=8$ head ($0.98\%$) and
the tightest but one under the token ratio --- so the ordering is a property of the
language \emph{and} the read-out rather than of the language alone. The full
per-language panel is Table~\ref{tab:fprlang}: every language with at least $100$
files, all four read-outs, at the deployment operating point. It carries the two
facts the five-language slice cannot: the heads disagree about which languages are
hard (C-family C, C\# and Dart are where the $K{=}16$ head concentrates its worst
rates, while the token ratio is loosest on PHP, Ruby and Objective-C), and every
read-out has a language where its rate is at least $3\times$ its aggregate ---
$3.29\%$ on Objective-C for the line head, $2.21\%$ on C for the $K{=}16$ head
against a $0.55\%$ aggregate, $3.53\%$ on Dart for the $K{=}8$ head. One caveat on
coverage: the block head's language feature covers $14$ languages and everything
else falls into an \emph{other} bucket, and the worst per-language rate at the
deployment operating point --- Dart at $K=16$ ($4.31\%$, $n=255$) --- sits in that
bucket, so it is a read-out-specific failure rather than a language the corpus
never showed the head.

\begin{table}[tbp]
\small
\caption{False-positive rate (\%) per language at the deployment operating point
(the $99$th-percentile development threshold of Table~\ref{tab:fpr}, transported
unchanged), on the same $19{,}445$ pre-2022 repository files. Every language with
at least $100$ files is tabulated, in descending order of file count; languages
below that cutoff are pooled into \emph{other} ($279$ files). The block head's
language feature covers $14$ languages; Dart, Objective-C, Swift, Kotlin, Lua and
Rust --- and everything in \emph{other} --- fall outside it. Lower is better; a
rate of $0.00$ is an exact zero, not a missing entry.}
\label{tab:fprlang}
\begin{center}
\scriptsize
\setlength{\tabcolsep}{8pt}
\begin{tabular}{lrrrrr}
\toprule
Language & $n$ & Token ratio & Block $K{=}8$ & Block $K{=}16$ & Line head \\
\midrule
JavaScript   & 4{,}545 & 0.86 & 0.84 & 0.68 & 1.67 \\
Java         & 3{,}173 & 0.28 & 0.44 & 0.28 & 0.35 \\
Python       & 3{,}038 & 0.33 & 0.30 & 0.13 & 0.39 \\
C            & 1{,}721 & 0.41 & 1.51 & 2.21 & 0.58 \\
C\#          & 1{,}190 & 0.76 & 1.60 & 0.17 & 0.67 \\
PHP          & 1{,}125 & 1.42 & 0.44 & 0.00 & 0.71 \\
C++          & 861     & 0.70 & 0.35 & 0.35 & 0.58 \\
TypeScript   & 797     & 0.38 & 0.13 & 0.00 & 0.88 \\
Ruby         & 652     & 1.23 & 0.46 & 0.15 & 0.61 \\
Go           & 614     & 0.16 & 0.98 & 0.49 & 0.65 \\
Shell        & 382     & 0.26 & 0.00 & 0.00 & 0.00 \\
other        & 279     & 0.36 & 0.72 & 0.36 & 0.36 \\
Dart         & 255     & 0.00 & 3.53 & \textbf{4.31} & 1.18 \\
Objective-C  & 213     & 1.88 & 1.41 & 0.94 & 3.29 \\
Swift        & 173     & 1.73 & 0.00 & 0.00 & 0.58 \\
Kotlin       & 154     & 0.00 & 0.00 & 0.00 & 0.00 \\
Lua          & 140     & 0.00 & 0.71 & 0.71 & 0.71 \\
Rust         & 133     & 0.00 & 0.75 & 0.00 & 0.00 \\
\bottomrule
\end{tabular}
\end{center}
\end{table}

\begin{table}[tbp]
\small
\caption{False-positive rate (\%) on $19{,}445$ pre-2022 repository files. All
files are negatives, so only FPR is defined. Every threshold is calibrated on the
development split ($2{,}031$ human files) and transported unchanged: the upper
panel uses the $99$th percentile of the development human scores (a nominal $1\%$
FPR on development, transported here rather than read off the test \textsc{human}
cell as in Table~\ref{tab:main}), the lower panel the
accuracy-maximising threshold. ``Synthetic'' applies the same threshold to the
$\Nhuman$ human files of the internal test set and is the reference point for the
``all'' column. Lower is
better everywhere; per-language file counts are in
Appendix~\ref{app:tables}.}
\label{tab:fpr}
\begin{center}
\scriptsize
\setlength{\tabcolsep}{4.5pt}
\begin{tabular}{lrrrccccc}
\toprule
Read-out & threshold & synthetic & all & Python & Java & JavaScript & TypeScript & Go \\
\midrule
\multicolumn{9}{l}{\emph{$1\%$ FPR operating point (deployment)}} \\
Token ratio, file mean & $0.0116$ & 0.67 & 0.60 & 0.33 & 0.28 & 0.86 & 0.38 & 0.16 \\
\quad + block head, $K=8$ & $0.383$ & 0.63 & 0.72 & 0.30 & 0.44 & 0.84 & 0.13 & 0.98 \\
\quad + block head, $K=16$ & $-0.303$ & 0.35 & \textbf{0.55} & 0.13 & 0.28 & 0.68 & 0.00 & 0.49 \\
\quad + line head, file mean & $1.364$ & 0.87 & 0.81 & 0.39 & 0.35 & 1.67 & 0.88 & 0.65 \\
\midrule
\multicolumn{9}{l}{\emph{Accuracy-maximising threshold (reported for completeness)}} \\
Token ratio, file mean & $0.00081$ & 5.08 & 4.73 & 3.62 & 2.40 & 6.58 & 4.89 & 1.79 \\
\quad + block head, $K=8$ & $-1.434$ & 7.71 & 8.66 & 3.46 & 7.03 & 11.11 & 6.15 & 9.93 \\
\quad + block head, $K=16$ & $-2.107$ & 5.67 & 7.31 & 3.23 & 4.88 & 8.84 & 4.64 & 6.35 \\
\quad + line head, file mean & $-0.100$ & 11.29 & 9.99 & 7.04 & 8.70 & 14.30 & 11.79 & 6.35 \\
\bottomrule
\end{tabular}
\end{center}
\end{table}

\begin{table}[tbp]
\small
\caption{False-alarm rate (\%) at block and line granularity, on the same
$19{,}445$ pre-2022 repository files as Table~\ref{tab:fpr}. Every file is a
negative, so every block and every line inside it is a negative too and only FPR
is defined: $n$ is the number of blocks (or lines) scored, not the number of
files, and the rate is the share of those units above threshold. Thresholds are
the granularity-matched thresholds frozen on the development split and
transported unchanged --- the block-level threshold of
Table~\ref{tab:blockcell} ($0.3214$ at $K=8$, $0.2271$ at $K=16$) and the
line-level threshold of Table~\ref{tab:linecell} ($-0.272$) --- so they are
\emph{not} the file-level operating points of Table~\ref{tab:fpr} and the two
tables are not comparable row by row. The token ratio has no sub-file read-out
and is absent. $53{,}525$ lines are not covered by the line head and are
excluded from its denominator. Lower is better.}
\label{tab:fprgran}
\begin{center}
\scriptsize
\setlength{\tabcolsep}{6pt}
\begin{tabular}{lrrr}
\toprule
Read-out & threshold & $n$ & FPR (\%) \\
\midrule
Block head, $K=8$ (blocks)  & $0.3214$  & 259{,}682     & 9.90 \\
Block head, $K=16$ (blocks) & $0.2271$  & 134{,}703     & 8.89 \\
Line head (lines)           & $-0.272$  & 1{,}955{,}822 & 17.33 \\
\bottomrule
\end{tabular}
\end{center}
\end{table}

\subsection{Implementation details}
\label{app:impl}

\begin{table}[tbp]
\small
\caption{Internal dataset (\texttt{aicd\_0907}). Every split is a set of files
labelled human or AI-involved; mixed files additionally carry character-level AI
spans, from which both the line labels and the block labels are derived
(Section~\ref{sec:problem}). Pairs count as two files: train / part1 is $6{,}001$
pairs, dev / pure $2{,}031$ pairs, and train / part2 is $1{,}582$ mixed files
($843$ continuation and $739$ fill-in-the-middle) each
accompanied by its own human original. The test set applies one generation axis at
a time (\emph{cells}) to a fixed human base: $11{,}773$ positives over six
generation cells against one shared pool of $\Nhuman$ human negatives.}
\label{tab:data}
\begin{center}
\scriptsize
\begin{tabular}{llrl}
\toprule
Split & Content & \#files & Used for \\
\midrule
train / part1 & pure human/AI pairs        & 12{,}002 & the two LoRA adapters \\
train / part2 & mixed files + their human originals & 3{,}164 & the block head \\
dev / mixed   & mixed files                & 398 & $K$, head hyper-parameters \\
dev / pure    & pure human/AI pairs        & 4{,}062 & aggregation and threshold \\
test          & 6 generation cells + shared \textsc{human} pool & 14{,}314 & reporting only \\
\bottomrule
\end{tabular}
\end{center}
\end{table}

The scoring backbone is Qwen/Qwen2.5-Coder-1.5B, kept frozen
throughout. Each adapter is a rank-$16$ LoRA module with $\alpha=32$ and $18.5$M
trainable parameters, targeting $\texttt{q\_proj}$, $\texttt{k\_proj}$,
$\texttt{v\_proj}$, $\texttt{o\_proj}$, $\texttt{gate\_proj}$,
$\texttt{up\_proj}$ and $\texttt{down\_proj}$, and trained with AdamW at
learning rate $10^{-4}$ for one epoch on its own side of the pure-file split with
a plain causal language modeling objective, truncating each file to $2048$ tokens
(the $90$th percentile of the length distribution; $5$--$7\%$ of training files
are cut). Training and scoring run in fp16: the available accelerators are
Volta-class and have no bf16 hardware, so bf16 would fall back to an emulation
path. Batching is by token budget over length-sorted buckets, with gradient
checkpointing enabled and no sequence packing.

Scoring is windowed rather than truncated. The per-token profile pass uses a
$4096$-token window with $512$ tokens of carried context, which keeps files of any
length scoreable. Training truncates at $2048$ tokens while scoring reads
$4096$-token windows, so the positions and relative distances beyond $2048$ are
extrapolated at scoring time: the backbone's rotary embeddings are defined up to
$32{,}768$ positions, so this is a distribution shift rather than an out-of-range
error, and every scored token still carries at least $512$ tokens of real left
context. The logit computation is sliced so that the vocabulary-sized
intermediate never materializes for the whole window at once. The base-model pass
additionally emits entropy and rank, and a second pass over a short context window
supplies the short-context contrast features. Token scores are persisted at
float32, so changing $K$ does not require re-scoring.

The block head is a bidirectional GRU over the $88$-dimensional block feature
sequence, trained on GPU; mixed blocks are masked out of the loss. Its aggregator
and $K$ are chosen on the development split. Both heads use $2$ layers of hidden
size $64$, dropout $0.1$, Adam \citep{kingma2015adam} at $10^{-3}$, at most $60$ epochs with early
stopping on the development split after $8$ epochs without improvement, and seed
$20260907$; the block head batches $32$ files at a time while the line head
batches by a padded-token budget of $131{,}072$, since single files of several
thousand lines make a fixed file count overflow memory.

\paragraph{Head input features.} Table~\ref{tab:featlist} lists both heads'
inputs. The block head's $88$ columns form five families, named by the prefix
that the drop-family rows of Table~\ref{tab:ablation} remove.
$\log$-rank is $\log(1+\mathrm{rank})$: the raw rank is long-tailed to
$65{,}535$, and on a linear scale a few large values dominate the block mean.
Every signal statistic is given twice, once as a raw value and once as a
within-file $z$-score whose mean and standard deviation are computed after
$1\%$/$99\%$ winsorization, and the two are not redundant. With raw values
only, how predictable a file is overall becomes a confound that masks
between-block differences; with $z$-scores only, every block of a pure-AI file
sits near its own file mean and is hard to separate from a block of a pure human
file, while the file score is an aggregate of block logits. Removing either
family is detectable but small on the test cells ($-0.008$ and $-0.011$ AUROC,
Table~\ref{tab:ablation}), so this is a design argument, not a large effect.
Blocks containing no token at all are dropped rather than zero-filled, which
would enter the head as neutral evidence; for an out-of-range or empty
neighbour in the \texttt{ctx} family the neighbour value falls back to the block
itself, the difference is set to zero, and a miss flag is raised. The line head
takes the same token profile but only $16$ columns, and in particular no
positional, language or size feature.

\begin{table}[tbp]
\small
\centering
\caption{Head inputs. \emph{Block head}: $88$ columns per block of $K$ lines,
family prefixes as used by the drop-family rows of Table~\ref{tab:ablation}.
\emph{Line head}: $16$ columns per token. $s_{\cdot}$ are the likelihood ratios
of Section~\ref{sec:method:lr}, $H$ entropy, and $\mathrm{dctx}$ the long- minus
short-context NLL gap, negated so that larger is more AI-like.}
\label{tab:featlist}
\begin{tabular}{llr}
\toprule
Family & Columns & \# \\
\midrule
\multicolumn{3}{l}{\emph{Block head}} \\
\texttt{raw} & $\slr$: mean, sd, min, max, $q_{10,25,50,75,90}$, positive rate & 10 \\
 & $\sai$, $\shum$, $\mathrm{nll}_{\mathrm{base}}$: mean, sd, $q_{50}$ & 9 \\
 & $H_{\mathrm{base}}$: mean, $q_{50}$; $\log$-rank and top-1 rate: mean & 4 \\
 & $-\mathrm{dctx}$ for the base, AI and human passes: mean & 3 \\
\texttt{raw.size} & $\log$ token count, line coverage, tokens and characters per line, block lines & 5 \\
\texttt{z} & within-file $z$-score of the $26$ statistics above & 26 \\
\texttt{ctx} & offsets $\pm 1, \pm 2$: neighbour $\slr$ mean, difference, miss flag & 12 \\
\texttt{pos} & relative index, first/last flags, $\log$ block count & 4 \\
\texttt{lang} & one-hot over $14$ languages and other & 15 \\
\cmidrule(l){2-3}
 & total & 88 \\
\midrule
\multicolumn{3}{l}{\emph{Line head} (per token; no \texttt{pos}, \texttt{lang} or size column)} \\
raw & $\mathrm{nll}$ of the base, AI and human passes; $H_{\mathrm{base}}$; $\log$-rank; the three short-context $\mathrm{nll}$ & 8 \\
$z$ & within-file $z$-score of the same eight & 8 \\
\cmidrule(l){2-3}
 & total & 16 \\
\bottomrule
\end{tabular}
\end{table}

\paragraph{Feature definitions.} Let $T_b$ be the token set of block $b$, $B$ the
number of blocks and $\mathcal{B}_{\mathrm{val}}$ the blocks holding at least one
token. With $\mathrm{nll}_y(t) = -\log p_y(x_t \mid x_{<t})$ for
$y \in \{\theta, \mathrm{ai}, \mathrm{hum}\}$ and $\mathrm{nll}^{\mathrm{s}}_y(t)$
its short-context counterpart, the token signals behind both heads are
\begin{equation}
\begin{aligned}
{\slr}_t &= \mathrm{nll}_{\mathrm{hum}}(t) - \mathrm{nll}_{\mathrm{ai}}(t), &
{\sai}_t &= \mathrm{nll}_{\theta}(t) - \mathrm{nll}_{\mathrm{ai}}(t), \\
{\shum}_t &= \mathrm{nll}_{\mathrm{hum}}(t) - \mathrm{nll}_{\theta}(t), &
\mathrm{dctx}^{y}_{t} &= \mathrm{nll}_y(t) - \mathrm{nll}^{\mathrm{s}}_y(t),
\end{aligned}
\label{eq:featsignals}
\end{equation}
for $y \in \{\theta, \mathrm{ai}, \mathrm{hum}\}$, all oriented so that larger
means more AI-like: $\mathrm{dctx}$ is the context gain negated, since
machine-written tokens gain less from surrounding context.
The \texttt{raw} family reduces a signal $v$ over $T_b$ with
\begin{equation}
\begin{aligned}
\bar v_b &= \frac{1}{|T_b|}\sum_{t \in T_b} v_t, &
\mathrm{sd}(v_b) &= \Big(\frac{1}{|T_b|}\sum_{t \in T_b} (v_t - \bar v_b)^2\Big)^{1/2}, \\
& \min_{t \in T_b} v_t, \quad \max_{t \in T_b} v_t, \quad q_p(v_b),
\end{aligned}
\label{eq:blockstats}
\end{equation}
$q_p$ being the empirical $p$-quantile over $T_b$, $\mathrm{pos}(v_b)$ the share
of its tokens with $v_t > 0$, and $r_t$ the rank of $x_t$ under $p_{\theta}$;
Table~\ref{tab:featlist} names which of these each signal takes. The \texttt{z} family standardizes each of the $26$ raw statistics across
the file,
\begin{equation}
z(v_b) = \frac{\mathrm{clip}(v_b,\, w_{0.01},\, w_{0.99}) - \mu_v}{\max(\sigma_v, \epsilon)},
\qquad \epsilon = 10^{-6},
\label{eq:blockz}
\end{equation}
where $w_{0.01}$ and $w_{0.99}$ are the $1\%$ and $99\%$ quantiles of
$\{v_{b'}\}_{b' \in \mathcal{B}_{\mathrm{val}}}$ and $\mu_v$, $\sigma_v$ the mean
and standard deviation of those winsorized values, the numerator being
winsorized the same way; a constant column yields $0$ rather than an infinity.
The \texttt{raw.size} family adds
\begin{equation}
\log(1{+}|T_b|), \quad
\mathrm{cov}_b = \frac{|\{\ell \in b : \ell \text{ carries a token}\}|}{K}, \quad
\frac{|T_b|}{K}, \quad \frac{1}{K}\sum_{\ell \in b} c_\ell, \quad K,
\label{eq:blocksize}
\end{equation}
with $c_\ell$ the character length of line $\ell$ including its newline. For each
$\delta \in \{-2,-1,+1,+2\}$ the \texttt{ctx} family carries the neighbour mean
$\bar s_{b+\delta}$, the contrast $\bar s_b - \bar s_{b+\delta}$ and a miss flag;
when $b+\delta$ leaves $\{1,\dots,B\}$ or is empty, the neighbour value falls back
to $\bar s_b$, the contrast is $0$ and the flag is $1$. The \texttt{pos} family is
$\big((b-1)/\max(B-1,1),\; \mathbb{1}[b{=}1],\; \mathbb{1}[b{=}B],\;
\log(1{+}B)\big)$, and \texttt{lang} is a one-hot over the $14$ corpus languages
plus \emph{other}, identical for every block of a file.

The line head reads the same profile at token granularity. Its eight signals
$[\mathrm{nll}_{\theta}, \mathrm{nll}_{\mathrm{ai}}, \mathrm{nll}_{\mathrm{hum}},
H_{\theta}, \log(1{+}r),
\mathrm{nll}^{\mathrm{s}}_{\theta}, \mathrm{nll}^{\mathrm{s}}_{\mathrm{ai}},
\mathrm{nll}^{\mathrm{s}}_{\mathrm{hum}}]$ are each fed twice, as the raw value
and as the within-file $z$-score of Equation~\ref{eq:blockz} taken over all
tokens of the file rather than over blocks, which gives the $16$ columns of
Table~\ref{tab:featlist}. It adds no positional, language or size column, and its
only within-line aggregation is the mean of the token logits over the tokens
assigned to that line.

The two LoRA adapters use learning rate $10^{-4}$ and truncation length
$2048$ tokens; the block and line heads use Adam at learning rate $10^{-3}$.
The classification-head control is trained and scored with the same
$4096$-token window and $512$-token carried context as the windowed scorer (stride
$4096-512=3584$; its file score is the mean over window logits), so its context is
no longer shorter than ours, as stated in Section~\ref{sec:results}.

\subsection{Table conventions and protocol details}
\label{app:tables}

Unless a caption says otherwise: each cell reads AUROC / TPR@1\%FPR / macro F1, the
latter two in percent; on the internal test set the negative pool is the shared
\textsc{human} cell ($\Nhuman$ files) and the positive class is AI-involved with
mixed files included ($\rho \ge 0.05$); \emph{Avg.} or \emph{all cells} is the
unweighted mean over the six generation cells and \emph{mixed} the mean over
\textsc{compl.}, \textsc{refac.}\ and \textsc{fim}; chance is AUROC $0.500$ and TPR
$1\%$, while the macro F1 floor at the $1\%$ FPR operating point is the
all-negative predictor, which a score without signal approaches there; it depends
on class balance and is quoted per caption, and the all-positive predictor is not
used as a floor because its FPR is $100\%$; bold marks the best value in each column for each metric, and a dash
means ``not defined'', not ``not run''. The threshold behind both operating-point
metrics --- TPR@1\%FPR and, where it is reported at that point, macro F1 --- is the
$99$th percentile of that table's own negative pool, which on the internal test
set is the \textsc{human} cell of the test set and not a development quantity, and
it is recomputed inside each replicate wherever a bootstrap interval is given. The
ablation tables of Section~\ref{sec:ablation} and
Appendix~\ref{app:ablation-extra} inherit that convention; the tables whose
thresholds are instead frozen on the development split
(Tables~\ref{tab:blockcell}, \ref{tab:linecell}, \ref{tab:acccell},
\ref{tab:fprlang} and~\ref{tab:fpr}) say so in their captions.

\paragraph{Table~\ref{tab:main} (internal test set).} Table~\ref{tab:main-full}
below is the same comparison with all three metrics per cell; Table~\ref{tab:main}
in the main text carries its macro F1 rows and its \emph{Avg.} column unchanged.
Per-cell positive counts are
\textsc{base} $2{,}541$, \textsc{unseen} $2{,}541$, \textsc{rewrite} $1{,}996$,
\textsc{compl.} $1{,}496$, \textsc{refac.} $1{,}899$ and \textsc{fim} $1{,}300$.
The \textsc{unseen} cell splits over its four writer models as $634$
(Qwen3-Coder-Plus), $634$ (Doubao-Seed-2.1-Pro), $637$ (Kimi-K2.6) and $636$
(MiniMax-M2.7).
The aggregation is fixed on the development split and applied unchanged to test,
and the $1\%$ FPR point is read off the \textsc{human} cell itself; macro F1 uses
that same operating point. The TPR@1\%FPR average is tied at $48.4$ between $K=8$
and $K=16$, so both are bold. The line head is a bidirectional GRU over the token
signals emitting one score per line; its file score is the mean of its line
logits, which beats both a max and a learned attention aggregator on the held-out
development files. The \textsc{aigcode} row is scored on a stratified subsample of
$500$ files per cell ($3{,}500$ in total) rather than the full test set, and its
composite score is our reconstruction: the original paper publishes neither the
mask-weight function, nor a burstiness formula, nor the three weights. The
\textsc{droiddetect} row is a file-level ModernBERT classifier truncated to
$2{,}048$ tokens, not one of the chunk-level encoders it sits among.

\begin{table}[tbp]
\small
\caption{File-level detection on the internal test set, all three metrics per cell.
This is Table~\ref{tab:main} of the main text with AUROC and TPR@1\%FPR restored;
the macro F1 row of every cell and the \emph{Avg.} column are identical to it.
\tabconv{} The two operating-point metrics share the threshold read off the
\textsc{human} cell, and the all-negative predictor scores $33.3$--$39.8\%$ macro F1
across the cells. Group~C rows are cumulative, not independent detectors.}
\label{tab:main-full}
\centering
\scriptsize
\renewcommand{\arraystretch}{1.15}
\setlength{\tabcolsep}{7pt}
\begin{tabular}{llccccccc}
\toprule
& Method & \textsc{base} & \textsc{unseen} & \textsc{rewrite} & \textsc{compl.} & \textsc{refac.} & \textsc{fim} & Avg. \\
& & \multicolumn{7}{c}{\scriptsize each cell, top$\to$bottom: AUROC / TPR@1\%FPR / macro F1 (\%)} \\
\midrule
\multirow{6}{*}{A}
& log-rank & \shortstack{0.415\\0.2\\33.3} & \shortstack{0.416\\0.1\\33.2} & \shortstack{0.441\\0.1\\35.7} & \shortstack{0.584\\0.7\\39.1} & \shortstack{0.507\\0.3\\36.5} & \shortstack{0.566\\1.3\\40.9} & \shortstack{0.488\\0.4\\36.5} \\
& LRR & \shortstack{0.401\\0.2\\33.4} & \shortstack{0.415\\0.1\\33.2} & \shortstack{0.430\\0.1\\35.7} & \shortstack{0.569\\0.5\\38.9} & \shortstack{0.491\\0.2\\36.4} & \shortstack{0.577\\0.7\\40.3} & \shortstack{0.481\\0.3\\36.3} \\
& Fast-DetectGPT & \shortstack{0.376\\0.2\\33.3} & \shortstack{0.397\\0.2\\33.3} & \shortstack{0.365\\0.2\\35.9} & \shortstack{0.685\\3.3\\41.8} & \shortstack{0.500\\0.4\\36.6} & \shortstack{0.566\\0.5\\40.1} & \shortstack{0.481\\0.8\\36.8} \\
& Binoculars & \shortstack{0.434\\0.1\\33.2} & \shortstack{0.433\\0.1\\33.2} & \shortstack{0.431\\0.1\\35.7} & \shortstack{0.595\\0.7\\39.1} & \shortstack{0.502\\0.5\\36.7} & \shortstack{0.532\\0.8\\40.4} & \shortstack{0.488\\0.4\\36.4} \\
& DetectCodeGPT & \shortstack{0.466\\0.2\\33.3} & \shortstack{0.477\\0.0\\33.1} & \shortstack{0.486\\0.0\\35.7} & \shortstack{0.583\\0.7\\39.1} & \shortstack{0.530\\0.3\\36.5} & \shortstack{0.569\\1.0\\40.6} & \shortstack{0.518\\0.4\\36.4} \\
& AIGCode Detector & \shortstack{0.460\\1.0\\34.1} & \shortstack{0.463\\0.8\\33.8} & \shortstack{0.478\\0.8\\33.8} & \shortstack{0.483\\1.2\\34.3} & \shortstack{0.471\\0.2\\33.2} & \shortstack{0.493\\1.6\\34.7} & \shortstack{0.475\\0.9\\34.0} \\
\midrule
\multirow{6}{*}{B}
& CodeT5+ + MLP & \shortstack{0.911\\35.5\\63.6} & \shortstack{0.906\\32.9\\61.7} & \shortstack{0.888\\25.1\\58.2} & \shortstack{0.619\\2.0\\40.5} & \shortstack{0.748\\7.1\\43.4} & \shortstack{0.551\\1.5\\41.1} & \shortstack{0.771\\17.3\\51.4} \\
& CodeBERT & \shortstack{0.943\\48.9\\72.2} & \shortstack{0.909\\37.5\\64.9} & \shortstack{0.939\\40.9\\69.1} & \shortstack{0.618\\2.1\\40.6} & \shortstack{0.770\\15.0\\50.6} & \shortstack{0.566\\1.9\\41.6} & \shortstack{0.791\\24.4\\56.5} \\
& UniXcoder & \shortstack{0.963\\53.1\\74.7} & \shortstack{0.945\\43.0\\68.5} & \shortstack{0.962\\46.4\\72.5} & \shortstack{0.647\\3.3\\41.8} & \shortstack{0.821\\16.1\\51.5} & \shortstack{0.565\\2.0\\41.7} & \shortstack{0.817\\27.3\\58.4} \\
& ImBD & \shortstack{0.861\\26.3\\57.0} & \shortstack{0.879\\27.4\\57.8} & \shortstack{0.848\\18.6\\53.1} & \shortstack{0.524\\2.3\\40.8} & \shortstack{0.688\\4.4\\40.8} & \shortstack{0.537\\1.6\\41.3} & \shortstack{0.723\\13.4\\48.5} \\
& \shortstack[l]{\emph{Same backbone}\\\quad + LoRA cls.\ head} & \shortstack{0.957\\61.1\\79.3} & \shortstack{0.944\\48.0\\71.6} & \shortstack{0.952\\50.8\\75.0} & \shortstack{0.635\\3.7\\42.2} & \shortstack{\textbf{0.854}\\17.8\\52.9} & \shortstack{0.505\\1.7\\41.3} & \shortstack{0.808\\30.5\\60.4} \\
& DroidDetect & \shortstack{\textbf{0.965}\\67.6\\82.9} & \shortstack{0.947\\64.0\\80.9} & \shortstack{0.965\\57.2\\78.6} & \shortstack{0.580\\5.2\\43.7} & \shortstack{0.804\\20.2\\54.9} & \shortstack{0.548\\2.5\\42.2} & \shortstack{0.801\\36.1\\63.8} \\
\midrule
\multirow{4}{*}{C}
& \shortstack[l]{Ours: token ratio,\\file mean} & \shortstack{0.964\\\textbf{79.4}\\\textbf{89.1}} & \shortstack{0.969\\80.1\\89.4} & \shortstack{\textbf{0.972}\\\textbf{74.2}\\\textbf{87.4}} & \shortstack{0.711\\12.0\\50.0} & \shortstack{0.810\\31.3\\63.1} & \shortstack{0.601\\5.7\\45.3} & \shortstack{0.838\\47.1\\70.7} \\
& + block head, $K=8$ & \shortstack{0.940\\74.0\\86.3} & \shortstack{0.967\\82.6\\90.7} & \shortstack{0.924\\65.9\\83.3} & \shortstack{0.805\\\textbf{25.5}\\\textbf{60.8}} & \shortstack{0.806\\30.9\\62.9} & \shortstack{\textbf{0.672}\\\textbf{11.2}\\\textbf{50.4}} & \shortstack{0.852\\\textbf{48.4}\\\textbf{72.4}} \\
& + block head, $K=16$ & \shortstack{0.954\\75.5\\87.1} & \shortstack{\textbf{0.974}\\\textbf{84.2}\\\textbf{91.6}} & \shortstack{0.944\\66.7\\83.7} & \shortstack{0.789\\22.5\\58.5} & \shortstack{0.819\\\textbf{32.3}\\\textbf{63.8}} & \shortstack{0.638\\9.2\\48.6} & \shortstack{\textbf{0.853}\\\textbf{48.4}\\72.2} \\
& + line head, file mean & \shortstack{0.923\\62.0\\79.8} & \shortstack{0.949\\67.7\\82.9} & \shortstack{0.889\\46.3\\72.4} & \shortstack{\textbf{0.817}\\20.6\\57.1} & \shortstack{0.769\\20.1\\54.8} & \shortstack{0.667\\7.6\\47.1} & \shortstack{0.836\\37.4\\65.7} \\
\bottomrule
\end{tabular}
\end{table}

\paragraph{Table~\ref{tab:seglen} (segment length).} Segments and block labels
follow Section~\ref{sec:problem}: a segment is a maximal run of AI lines after
merging runs separated by at most one human line, and a block is positive if it
contains at least one AI line. Every
variant scores the same positive files against the shared \textsc{human} cell
($\Nhuman$ files), except \emph{oracle-pair}, which scores only the tokens inside
the true AI spans and substitutes the same line
range of each positive's paired human original for the negative, so that content
difficulty is controlled. The \emph{length-only} row is a logistic probe on line
count and character count, trained on train split part~1. Positive and negative
blocks are both drawn from the
same mixed files. The file-level reading aggregates the same block logits by their
mean against the shared human pool.

\paragraph{Table~\ref{tab:ws} (formatting).} The whitespace probe uses nine
features: indentation mean, std, max and number of distinct levels; tab-indented
line ratio; blank-line ratio; longest blank run; trailing-whitespace line ratio;
and line count. The length probe uses line count and character count. Both are
trained on train split part~1. The probes are linear; tree ensembles can extract
somewhat more from the same features.

\paragraph{Tables~\ref{tab:fpr} and~\ref{tab:fprlang} (real human code).} The five
per-language columns of Table~\ref{tab:fpr} are the five largest language slices of
the corpus; Table~\ref{tab:fprlang} extends them to every language with at least
$100$ files, and its $n$ column is the per-language file count. The two tables come
from the same run and the same thresholds.

\paragraph{Table~\ref{tab:external-main} (external transfer).}
Chunk scores (encoders) and block probabilities (our block heads) are aggregated to
a file score by the aggregator selected on the corresponding development split:
logit-mean for both $K$ of the block heads, and mean for P1 against max for P2 on
DroidCollection. The line head's file score is the mean of its per-line logits,
which beats both a max and a learned attention aggregator on held-out development
files. Chance for TPR@1\%FPR is $0.010$. Thresholds follow what each benchmark
ships. TPR@1\%FPR always reads the $99$th percentile of the benchmark's own
negatives. On AICD-Bench T1 and T3, which ship no development split, macro F1 is
read at that same threshold, so those four F1 columns are calibrated on the test
negatives and are therefore optimistic; on DroidCollection, which does ship a development split, the macro F1
threshold is the value that maximizes pooled macro F1 there and is transported to
test unchanged. The \textsc{aigcode} row is zero-shot and
therefore has no P2; \textsc{droiddetect} is a file-level classifier truncated to
$2{,}048$ tokens. ImBD scores the conditional-probability curvature of an
SPO-aligned model over the full file, truncated to $8{,}192$ tokens; P1 transports
the internally aligned model unchanged. Its P2 needs one caveat. SPO is a DPO stage
over \emph{content-equivalent} pairs --- the same file written by a human and
rewritten by a model --- and no external benchmark provides them: on all three, the
machine-written and human sides are disjoint file sets with no shared identifier. We
therefore pair within each benchmark by bucketing on (language, source) and matching
on token length inside each bucket, which aligns the carrier distribution but not the
content, and train on a quarter of the pairs ($3{,}997$ / $5{,}077$ / $4{,}731$ for
T1 / T3 / DroidCollection) so that the SPO budget stays comparable to the $5{,}873$
pairs used internally. Held-out DPO accuracy reaches $1.000$ / $0.948$ / $0.943$, so
the alignment fits its own objective; what it aligns to is partly corpus identity
rather than style. Two consequences are visible in the table. On T3 adaptation
\emph{inverts} the ranking ($0.399$, italicized), the only sub-chance P2 entry
anywhere; and on all three the direction check on held-out pairs comes out negative,
i.e.\ the aligned model assigns \emph{lower} curvature to machine code, so the sign
is flipped before scoring. Read the ImBD P2 column as the cost of transplanting a
protocol whose data assumption the target corpus does not satisfy, not as a ceiling
on the method.

\paragraph{Table~\ref{tab:external-hca} (HybridCodeAuthorship).} The line level
carries $229{,}158$ attribution runs for our methods and $215{,}254$ for Group~A,
which drops runs the scoring tokenizer reduces to zero tokens; Group~B scores
$215{,}422$ runs, and $215{,}258$ for ImBD, which skips runs its tokenizer
empties. The Block columns carry $350{,}755$ ($K{=}8$) and $178{,}050$ ($K{=}16$)
blocks, scored for every row; each method is scored on the unit of the column it
fills: a Line cell scores one attribution run and takes that line's own label, a
Block cell scores the $K$-line block as a single text and counts it positive if the
block contains at least one AI line, and a File cell scores the whole file, so
\textsc{aigcode} (perturbation sampling) and \textsc{droiddetect} fill all three.
DetectCodeGPT was not run. HCA ships no training or development split, so no
threshold in this table is calibrated on HCA itself. Our rows take their line- and
block-level thresholds from the internal development split, frozen there and
transported unchanged --- the line threshold is the $-0.272$ of
Table~\ref{tab:linecell}, the block thresholds come from the internal $K{=}16$
development blocks --- which is a granularity mismatch we record rather than
repair, since those values were selected on internal lines and blocks and not on
HCA attribution runs. The File column instead reads the $99$th percentile of HCA's
own negatives, so its $F1$ is calibrated on the test negatives in the same way as
the internal macro F1 columns. Group~B baselines are internal
fine-tunes transported unchanged: encoders score tree-sitter chunks (the internal
$\le 512$-token splitting) aggregated by mean, the LoRA head scores the
$4096$-token windows aggregated by mean, and ImBD scores the
conditional-probability curvature of its
SPO-aligned model over the full file; at the line level every Group~B row scores
each attribution run directly with its own file-level input recipe. At the file
level the zero-shot direction inverts (AUROC far below $0.500$): the purely human
files average as \emph{more} machine-like than the co-written ones, i.e.\ the
detectors rank boilerplate-heavy human code as generated. At the file level the
Group~B rows sit at chance-to-weak ($0.472$--$0.594$), apart from DroidDetect at
$0.712$; at the line level CodeBERT
is the best of them at $0.666$, below our block head at $K=8$ ($0.685$), while the
LoRA head \emph{inverts} to $0.337$ and ImBD reaches only $0.536$. The two are not
at the same granularity, so the comparison places our block read-out at the upper
end of the baselines rather than clear of them.

\subsection{External benchmarks: composition and per-slice results}
\label{app:external}

Table~\ref{tab:external-data} fixes what each external benchmark contributes, and
Table~\ref{tab:external-hca} reports HybridCodeAuthorship at all four
granularities.

\begin{table}[tbp]
\small
\caption{External benchmarks used in Section~\ref{sec:results:external}.
\emph{Granularity} is the unit at which the benchmark carries ground truth (we
always report a file-level decision, and additionally match the native
granularity where one exists). \emph{\#eval} is the number of files we score
after subsampling, with the negative count in parentheses. Without a train split
only P1 is defined. \emph{Published} is the best number reported by the
benchmark's own paper together with its metric, which is in general neither our
metric nor our protocol; it is included only as a same-protocol reference for our
P2 column, and \tbd{} means the benchmark reports no comparable number --- as
opposed to the ``--'' that Section~\ref{sec:setup:metrics} reserves for ``not
defined''.}
\label{tab:external-data}
\begin{center}
\scriptsize
\setlength{\tabcolsep}{3pt}
\begin{tabular}{llllll}
\toprule
Benchmark & Granularity & Native label & \#eval (neg.) & Train split & Published (metric) \\
\midrule
AICD-Bench T1 & file & binary   & 20{,}000 (15{,}519) & yes & 62.31 (macro-F1, zs-LLM) \\
AICD-Bench T3 & file & 4-class & 20{,}000 (13{,}447) & yes & 61.65 (macro-F1, FT) \\
DroidCollection & file        & 4-class & 105{,}826 (49{,}932) & yes & 97.00 (domain-avg wF1, 2-cls, FT) \\
HCA           & line, chunk & per-line binary & 10{,}488 (1{,}904) & no  & 56.0 (F1, line, zs) \\
\bottomrule
\end{tabular}
\end{center}
The four DroidCollection classes are \emph{human}, \emph{machine-generated}
($26{,}524$), \emph{machine-refined} ($13{,}230$) and \emph{adversarial}
($16{,}140$); we map the latter three to $y=1$ and evaluate them both pooled and
separately. The published entries read as follows. AICD-Bench T1's $62.31$ is a
zero-shot chain-of-thought LLM (Gemini-2.5-Flash) on a $5\%$ stratified
subsample; the best detector \emph{trained on the benchmark's own split} is
DeBERTa at $34.13$, below the random baseline of $45.73$. T3's $61.65$
(ModernBERT) and DroidCollection's $97.00$ (DroidDetect-Large, binary, domain-averaged) were both
obtained by training on the benchmark's own training split --- the same protocol
as our P2, but in macro/weighted F1 rather than AUROC.
\end{table}

\begin{table}[tbp]
\small
\caption{Full Group~A (zero-shot) transfer results, summarized by the
DetectCodeGPT row of Table~\ref{tab:external-main}. Group~A trains nothing, so only
the transfer protocol P1 is defined. Each cell is AUROC / TPR@1\%FPR / macro F1
(\%).}
\label{tab:external-zs}
\begin{center}
\scriptsize
\setlength{\tabcolsep}{3pt}
\begin{tabular}{lccc}
\toprule
Method & AICD-Bench T1 & AICD-Bench T3 & DroidCollection \\
\midrule
LogRank          & 0.665 / 5.5 / 48.8  & 0.626 / 0.0 / 40.0 & 0.626 / 3.3 / 35.4 \\
LRR              & 0.663 / 2.9 / 46.3  & 0.656 / 1.3 / 41.4 & 0.628 / 2.0 / 34.0 \\
Fast-DetectGPT   & 0.637 / 10.4 / 53.2 & 0.653 / 6.1 / 46.1 & 0.604 / 6.5 / 38.6 \\
Binoculars       & 0.529 / 1.0 / 44.5  & 0.406 / 0.0 / 40.0 & 0.497 / 0.4 / 32.3 \\
DetectCodeGPT    & 0.703 / 6.5 / 49.7  & 0.666 / 0.2 / 40.2 & 0.669 / 3.0 / 35.1 \\
AIGCode Detector & 0.505 / 1.8 / 45.2  & 0.438 / 1.1 / 40.7 & 0.510 / 2.7 / 34.8 \\
\bottomrule
\end{tabular}
\end{center}
\end{table}

\begin{table}[tbp]
\small
\caption{Full version of Table~\ref{tab:external-hca}: HybridCodeAuthorship
\citep{hybridcodeauthorship} under P1 with every row restored, i.e.\ the zero-shot
statistics and the CodeT5+ + MLP baseline that the main-text table omits.
\tabconv{} F1 here is plain, not macro. Every entry shared with
Table~\ref{tab:external-hca} is identical to it.}
\label{tab:external-hca-full}
\begin{center}
\scriptsize
\setlength{\tabcolsep}{3pt}
\renewcommand{\arraystretch}{0.85}
\begin{tabular}{llcccc}
\toprule
& \multirow{2}{*}{Method} & \multicolumn{4}{c}{HCA, P1 (zero adaptation)} \\
\cmidrule(lr){3-6}
& & Line & Block $K{=}8$ & Block $K{=}16$ & File \\
& & \multicolumn{4}{c}{\scriptsize AUROC / TPR@1\%FPR / F1 (\%)} \\
\midrule
\multirow{6}{*}{A}
& LogRank                                     & 0.509 / 0.6 / 1.1 & 0.592 / 0.3 / 0.6 & 0.587 / 0.3 / 0.6 & 0.007 / 0.0 / 0.0 \\
& LRR                 & 0.491 / 0.3 / 0.6 & 0.583 / 1.2 / 2.3 & 0.581 / 1.3 / 2.5 & 0.050 / 1.1 / 2.2 \\
& Fast-DetectGPT & 0.495 / 1.2 / 2.2 & 0.568 / 1.8 / 3.5 & 0.568 / 1.9 / 3.7 & 0.499 / 0.0 / 0.0 \\
& Binoculars       & 0.474 / 0.3 / 0.6 & 0.538 / 0.2 / 0.4 & 0.523 / 0.2 / 0.3 & 0.149 / 0.0 / 0.0 \\
& Binoculars (reverse)                        & 0.550 / 0.9 / 1.8 & 0.551 / 0.4 / 0.8 & 0.537 / 0.2 / 0.5 & 0.124 / 0.0 / 0.0 \\
& AIGCode Detector      & 0.646 / 1.4 / 2.7 & 0.513 / 2.2 / 4.1 & 0.501 / 1.7 / 3.2 & 0.496 / 1.4 / 16.7 \\
\midrule
\multirow{6}{*}{B}
& CodeT5+ + MLP                          & 0.572 / 2.8 / 5.5 & 0.566 / 2.3 / 4.3 & 0.586 / 2.6 / 5.0 & 0.570 / 2.1 / 4.1 \\
& CodeBERT                               & \textbf{0.666} / 6.1 / 11.4 & 0.577 / 4.8 / 9.0 & 0.589 / 5.4 / 10.1 & 0.594 / 3.3 / 6.3 \\
& UniXcoder                              & 0.581 / 3.6 / 6.8 & 0.562 / 4.8 / 9.0 & 0.570 / 5.2 / 9.8 & 0.580 / 2.4 / 4.6 \\
& \emph{Same backbone} + LoRA cls.\ head      & \emph{0.337} / 0.6 / 1.1 & 0.532 / 1.3 / 2.5 & 0.534 / 1.4 / 2.7 & 0.476 / 1.2 / 2.4 \\
& ImBD & 0.536 / 1.3 / 2.5 & 0.531 / 2.6 / 5.0 & 0.553 / 3.0 / 5.7 & 0.472 / 2.2 / 4.3 \\
& DroidDetect & 0.586 / 2.6 / 5.0 & 0.682 / 6.6 / 12.2 & \textbf{0.685} / 9.3 / 16.8 & 0.712 / 11.3 / 26.7 \\
\midrule
\multirow{4}{*}{C}
& Ours: token ratio, line mean           & 0.528 / 0.9 / 53.5 & 0.637 / 6.9 / 49.5 & 0.628 / 9.2 / \textbf{52.1} & 0.638 / 12.8 / 22.7 \\
& \quad + block head, $K=8$              & -- & \textbf{0.685} / \textbf{14.0} / \textbf{51.5} & -- & 0.690 / 15.3 / 26.5 \\
& \quad + block head, $K=16$             & -- & -- & 0.627 / \textbf{11.0} / 50.4 & 0.667 / 14.4 / 25.2 \\
& \quad + line head, run mean            & 0.589 / \textbf{10.6} / \textbf{62.5} & -- & -- & \textbf{0.786} / \textbf{19.1} / \textbf{32.0} \\
\bottomrule
\end{tabular}
\end{center}
\end{table}

Two entries in Group~C of Table~\ref{tab:external-hca} need their own reading.
First, the line column: the line head is the strongest of our read-outs there
($0.589$ / $10.6$ / $62.5$ against $0.528$ / $0.9$ / $53.5$ for the token ratio,
the only other read-out defined at that granularity), and its $F1$ is an order of
magnitude above every baseline's. Second, the file column:
the line head is the strongest row there ($0.786$ against $0.690$ / $0.667$ for
the block heads, $0.638$ for the token ratio and $0.712$ for DroidDetect, the
strongest baseline), so the usable margin on HCA comes from aggregating per-line
evidence over the file rather than from reading it line by line. The block columns
are read the same way as the line column: at a median span of two lines almost
every $K$-line block contains both authors, so a block label that fires on a
single AI line is close to a file-level label smeared over blocks, and the block
AUROC drops from $0.685$ at $K=8$ to $0.627$ at $K=16$ as the blocks get coarser.

\end{document}